\documentstyle[12pt,epsf]{article}

\newcommand{\eq}[1]{(\ref{#1})}
\newcommand{\ab}{( \alpha - \beta )} 
\newcommand{\nn}{\nonumber}
\newcommand{\fr}{\frac}
\newcommand{\mh}{m_{h^0}^2}
\newcommand{\mH}{m_{H^0}^2}
\newcommand{\mg}{m_{H^{\pm}}^2}
\newcommand{\ma}{m_{A^0}^2}
\newcommand{\mw}{m_W^2}
\newcommand{\mz}{m_Z^2}
\newcommand{\mb}{m_b^2}
\newcommand{\mt}{m_t^2}
\newcommand{\pw}{p_W^{}}
\newcommand{\pz}{p_Z^{}}
\newcommand{\pg}{p_\gamma^{}}

\newcommand{\pwsq}{p_W^2}
\newcommand{\pzsq}{p_Z^2}
\newcommand{\pgsq}{p_\gamma^2}

\newcommand{\al}{\alpha}
\newcommand{\be}{\beta}
\newcommand{\sn}{\sin}
\newcommand{\cs}{\cos}

\newcommand{\tnb}{\tan \beta}
\newcommand{\ctb}{\cot \beta}
\newcommand{\snab}{\sin (\alpha - \beta)}
\newcommand{\csab}{\cos (\alpha - \beta)}

\newcommand{\lt}{\left}
\newcommand{\rt}{\right}

\newcommand{\gmn}{g_{\mu\nu}}
\newcommand{\gsim}{\mbox{ \raisebox{-1.0ex}{$\stackrel{\textstyle >}
{\textstyle \sim}$ }}}
\newcommand{\lsim}{\mbox{ \raisebox{-1.0ex}{$\stackrel{\textstyle <}
{\textstyle \sim}$ }}}
\def\Journal#1#2#3#4{{#1} {\bf #2} (#4) #3}

\def\NPB{{\em Nucl. Phys.} B}
\def\PLB{{\em Phys. Lett.}  B}
\def\PRL{\em Phys. Rev. Lett.}
\def\PRD{{\em Phys. Rev.} D}

\def\EPC{{\em Eur. Phys. J.} C}
\def\PTP{\em Prog.~Theor.~Phys.}
\def\MPLA{{\em Mod. Phys. Lett.} A}
\def\IJMPA{{\em Int. Mod. Phys.} A}
\def\JHEP{\em J. High Ener. Phys.}
\def\ibid{\it ibid.}

\begin{document}
\topmargin 0pt
\oddsidemargin 1mm
\begin{titlepage}
\begin{flushright}
hep-ph/9911541
\end{flushright}

\setcounter{page}{0}
\vspace{10 mm}
\begin{center}
{\Large \bf    Possible enhancement of the
     $e^+ e^- \rightarrow H^\pm W^\mp$ cross section 
            in the two-Higgs-doublet model  }
\end{center} 
\vspace{15 mm}
\begin{center}
{\large \bf  Shinya Kanemura }\\    
\vspace{2mm}
{\em Institut f\"{u}r Theoretishe Physik der Universit\"{a}t Karlsruhe \\
     D-76128 Karlsruhe, Germany
  \footnote{e-mail: kanemu@particle.physik.uni-karlsruhe.de}}\\
\end{center}

\vspace{15 mm}

\begin{abstract}
The production process of the charged Higgs-boson associated with a 
$W$ boson at electron-positron colliders is discussed in the 
two-Higgs-Doublet Model (2HDM) and in the Minimal Supersymmetric Standard 
Model (MSSM). The process is induced at one-loop level in these models.
We examine how much the cross section can be enhanced by  
quark- and Higgs-loop effects. In the non-SUSY 2HDM, in addition to large 
top-bottom ($t$-$b$) loop effects for small $\tan \beta$ 
($\ll \sqrt{m_t/m_b}$), the Higgs-loop diagrams can contribute to the 
cross section to some extent for moderate $\tan \beta$ values.
For larger $\tan \beta$ ($\gg \sqrt{m_t/m_b}$), such enhancement by the 
Higgs non-decoupling effects is bounded by the requirement for validity 
of perturbation theory. In the MSSM with heavy super-partner particles,  
only the $t$-$b$ loops enhance the cross section  
while Higgs-loop effects are very small.\\

\vspace{1cm}
\end{abstract}
\vspace{1cm}
\begin{center}
{\large {\em To appear in European Physical Journal {\bf C}}}  
\end{center}

\end{titlepage}

\section{Introduction}

\hspace*{18pt}
The Higgs sector has not yet been confirmed experimentally. In near future  
a neutral Higgs boson may be discovered at Tevatron II or LHC, by which the 
standard picture of particle physics may be completed. 
The exploration of additional Higgs bosons will be then very important 
in order to confirm extended Higgs sectors from the minimal Higgs sector 
in the Standard Model (SM).  Actually various theoretical insights suggest 
such extensions; the supersymmetry (SUSY), extra CP-violating phases, 
a source of neutrino masses, a remedy for the strong CP problem and so on.   
Most of the extended Higgs models include charged and CP-odd Higgs bosons. 
Therefore discovery of a charged Higgs boson, $H^\pm$, or a CP-odd Higgs 
boson, $A^0$ will confirm extended versions of the Higgs sector directly. 
At LHC, search of these extra Higgs bosons is also one of the most important 
tasks. In addition, considerable precision measurement of high energy 
phenomena may be possible at future linear colliders (LC's) 
such as JLC, NLC and TESLA\cite{lc}. 

In this paper, we discuss the charged-Higgs-boson production process 
associated with a $W$ boson at LC's,  $e^+e^- \rightarrow H^\pm W^\mp$, 
in the two-Higgs-Doublet Model (2HDM) including the Minimal Supersymmetric 
Standard Model (MSSM) with super-heavy super-partner particles.  
By neglecting the electron mass the process disappears at tree level 
because of no tree $H^\pm W^\mp V$ couplings ($V = \gamma$ and $Z^0$) 
in these models.  
Since these couplings occur at one-loop level\cite{mp,cphi,kanemu},  
the process $e^+e^- \rightarrow H^\pm W^\mp$ is induced at this level.   
At LC's, one of the main processes for charged Higgs search is
the $H^\pm$-pair production\cite{eehh}, whose cross section may be large 
enough to be detected if $H^\pm$ is much lighter than the threshold 
$\sqrt{s}/2$. The process rapidly reduces for heavier $H^\pm$ even if 
the mass is below the threshold. In this case   
$e^+e^- \rightarrow H^\pm W^\mp$ becomes important as a complementary 
process if its cross section can be large enough to be detected. 
Our question here is how much this loop-induced process can grow in 
the non-SUSY 2HDM as well as in the MSSM.  

Magnitude of the cross section for $e^+e^- \rightarrow H^\pm W^\mp$ 
directly shows dynamics of particles in the loop because 
there is no tree-level contribution. We here consider one-loop 
contributions of quarks, gauge bosons and Higgs bosons.   
In particular, the top-bottom ($t$-$b$) loop effects are expected to be 
sizable, because the Yukawa-coupling constants are proportional to 
quark masses so that the decoupling theorem by Appelquist and 
Carazzone\cite{dec} 
is not applied to this case. The naive power-counting argument shows that 
quadratic quark-mass terms appear in the amplitude with a longitudinally 
polarized $W$ boson. Therefore the $t$-$b$ loops can greatly 
contribute to the cross section depending on $\tan\be$. 
In the non-SUSY 2HDM, the Higgs-loop contributions can also be large  
when the Higgs self-coupling constants are proportional to the Higgs 
boson masses. 
Effects of the heavy Higgs bosons in the loop then do not decouple in the 
large mass limit. Instead, the quadratic mass terms of these Higgs bosons 
can appear in the amplitude\cite{kanemu,kko,nondec}, so that larger 
Higgs-loop effects are expected for heavier Higgs bosons in the loop.  
In contrast, if masses of the extra Higgs bosons are determined mainly by 
an independent scale of the vacuum expectation value ($\sim 246$ GeV), the 
Higgs-loop contributions tend to decouple for large extra-Higgs-boson masses. 
The MSSM Higgs sector corresponds to this case, so its loop-effects cannot  
be so large. 
The main purpose of this paper is to confirm above discussion analytically 
and numerically and to see the possible enhancement of the cross section by 
these non-decoupling effects under the requirement for validity of 
perturbation theory\cite{uni,tri,tanb,kko}.
The information from available experimental data such as the $\rho$ parameter 
constraint\cite{gr} and the $b \rightarrow s \gamma$ 
results\cite{bsg_ex,bsg_mg} are also taken into account.    

We find that the cross section can be quite large for small $\tan\be$ 
($\ll \sqrt{m_t/m_b}$) by  the $t$-$b$ loop effects.  
In addition, in the non-SUSY 2HDM,  
the cross section can grow to some extent by the Higgs non-decoupling 
effects for moderate values of $\tan\be$. 
For larger $\tan\be$ ($\gg \sqrt{m_t/m_b}$) such enhancement 
by the Higgs-loop effects is strongly bounded by the condition 
for the perturbation, and the cross section becomes smaller. 
 
In Sec 2, the 2HDM is reviewed briefly to fix our notation. 
The calculation of the cross section is explained in Sec 3. 
After some analytic discussion on the amplitude in Sec 4, 
we present our numerical results in Sec 5. Conclusion is given in Sec 6.
Details of the analytic results of the calculation are shown in Appendix.

\section{The 2HDM}
 
\hspace*{18pt}
The 2HDM with a softly-broken discrete symmetry under the transformation  
$\Phi_1 \rightarrow \Phi_1$, $\Phi_2 \rightarrow - \Phi_2$ is assumed. 
The Higgs sector is given by
\begin{eqnarray}
  {\cal L}_{\rm THDM}^{int} 
                       & = &  \mu_1^2 \left| \Phi_1 \right|^2 + 
                              \mu_2^2 \left| \Phi_2 \right|^2 + 
                             \lt\{ \mu_{3}^2 
                                  \lt( \Phi_1^{\dagger} \Phi_2 \rt) 
                                     + {\rm \,h.c. \,} \rt\}     \nn  \\
                       &   & - \lambda_1 \left| \Phi_1 \right|^4
                             - \lambda_2 \left| \Phi_2 \right|^4
                             - \lambda_3 \left| \Phi_1 \right|^2 
                                \left| \Phi_2 \right|^2  - \lambda_4 
                                     \lt( {\rm Re }\Phi_1^{\dagger}
                                                  \Phi_2 \rt)^2
                             - \lambda_5 \lt( {\rm Im }\Phi_1^{\dagger}
                                                  \Phi_2 \rt)^2. 
\label{int}
\end{eqnarray}
This potential includes the MSSM Higgs sector as a special case.
We here neglect all the CP-violating phases just for simplicity and 
all the coupling constants and masses are then real in Eq.~\eq{int}.
From the doublets $\Phi_1$ and $\Phi_2$ ($\langle \Phi_i \rangle \equiv 
v_i/\sqrt{2}$ and $\sqrt{v_1^2 + v_2^2} \sim 246$ GeV), five massive 
eigenstates as well as three Nambu-Goldstone modes ($w^\pm$ and $z^0$) 
are obtained; that is, two CP-even neutral bosons $h^0$ and $H^0$ 
diagonalized by the mixing angle $\al$, one pair of the charged Higgs 
boson $H^{\pm}$, and one CP-odd neutral Higgs boson $A^0$, where  
$h^0$ is lighter than $H^0$.
In addition to the four mass parameters $m_{h^0}$, $m_{H^0}$, $m_{H^\pm}$ 
and $m_{A^0}$, we have two mixing angles $\alpha$ and $\beta$ 
($\tan \beta = v_2/v_1$) and one free dimension-full parameter $M$ 
corresponding to the soft-breaking mass 
($M^2 \equiv \mu_3^2/(\sin \beta \cos \beta)$). 
Tree-level relations among the coupling constants and the masses are 
then given by\cite{kanemu}
  \begin{eqnarray}
  \lambda_1 
&=& \fr{1}{2 v^2 \cs^2 \be} 
 (\cs^2 \al \;\mH + \sn^2 \al \;\mh - \sin^2 \be \;M^2), \label{qu1}\\
  \lambda_2 
&=& \fr{1}{2 v^2 \sn^2 \be} 
 (\sn^2 \al \;\mH + \cs^2 \al \;\mh - \cos^2 \be \;M^2), \label{qu2} \\
  \lambda_3 
&=& \fr{\sn 2\al}{v^2 \sn 2\be} (\mH - \mh) + \fr{2 \mg}{v^2}
      - \fr{1}{v^2} \;M^2 ,\label{qu3} \\
  \lambda_4 
&=& - \fr{2 \mg}{v^2} + \fr{2}{v^2} \; M^2, \label{qu4} \\
  \lambda_5 
&=& \fr{2}{v^2} (\ma - \mg).
\label{mass}
\end{eqnarray}

As for the Yukawa interaction, two kinds of couplings are possible in our  
model: we call them Model I and Model II according to Ref.~\cite{hhg}. 
The Yukawa interaction with respect to the charged-Higgs boson is expressed 
by
\begin{eqnarray}
  {\cal L}_{Htb} = \overline{b} 
      \lt\{  \frac{y_b}{2}  \tan \be (1 - \gamma_5) 
           + \frac{y_t}{2}  \cot \be (1 + \gamma_5) \rt\} t H^-  
       +  {\rm h.c.} \,,  \label{yukawa}
\end{eqnarray}
where 
\begin{eqnarray}
&&       y_b = \frac{\sqrt{2} m_b}{v} \cot\be, \;\;\; 
         y_t = \frac{\sqrt{2} m_t}{v} \cot\be, \;\;\; 
        ({\rm Model}\; \rm{I}), \\
{\rm or\;\;}&&       y_b = \frac{\sqrt{2} m_b}{v} \tan\be, \;\;\;
         y_t = \frac{\sqrt{2} m_t}{v} \cot\be, \;\;\;  
        ({\rm Model}\; \rm{II}). 
\end{eqnarray}
Here Model II corresponds to the MSSM Yukawa-interaction.

\section{The calculation for $e^+ e^- \rightarrow H^- W^+$}

\hspace*{18pt}
We consider the process $e^-(\tau, k) + e^+(-\tau, \overline{k}) \rightarrow 
 H^-(p) + W^+(\overline{p},\overline{\lambda})$, 
where $\tau = \pm 1$ and $\overline{\lambda}= 0, \pm 1$ are helicities of the 
electron and the $W^+$ boson; $k$ and $\overline{k}$ are incoming momenta of 
the electron and the positron, while  
$p$ and $\overline{p}$ are outgoing momenta of $H^-$ and $W^+$, 
respectively. The helicity amplitude may be written by   
\begin{eqnarray}
  {\cal M}(k,\overline{k},\tau; p, \overline{p}, \overline{\lambda}) 
 = \sum_{i = 1}^3 F_{i,\tau}(s,t) \; K_{i,\tau} 
    (k,\overline{k},\tau; p, \overline{p}, \overline{\lambda}),
\end{eqnarray}
where the form factors $F_{i,\tau}(s,t)$ include all the dynamics that 
depends on the model. The kinematical factors are expressed by 
\begin{eqnarray}
   K_{i,\tau} 
    (k,\overline{k},\tau; p, \overline{p}, \overline{\lambda})
 = j_{\mu}(k,\overline{k},\tau) T^{\mu\beta}_i 
  \epsilon_\beta(\overline{p},\overline{\lambda})^\ast, 
\end{eqnarray}
where $j_\mu(k,\overline{k},\tau)$ is the electron current and 
$\epsilon_\beta(\overline{p},\overline{\lambda})^\ast$ 
is the polarization vector of the $W$ boson. 
The basis tensors $T^{\mu\beta}_i$ are defined by  
\begin{eqnarray}
  T^{\mu\beta}_1 &=& g^{\mu\beta}, \\ 
  T^{\mu\beta}_2 &=&   \fr{1}{m_W^2} P^\mu P^\beta,            \\
  T^{\mu\beta}_3 &=&   \fr{i}{m_W^2} \epsilon^{\mu\beta\rho\sigma} 
                         P_\rho q_\sigma, 
\end{eqnarray}
where $P^\mu \equiv p^\mu - \overline{p}^\mu$,  
      $q^\mu \equiv p^\mu + \overline{p}^\mu = k^\mu + \overline{k}^\mu$ and 
$\epsilon^{0123} = -1$.  
In Table 1, the explicit expressions for each $K_{i,\tau}$ in the 
center-of-mass frame are listed by using $\beta_{HW}$ and the scattering 
angle $\Theta$ 
\begin{eqnarray}
  \beta_{HW} &=& \sqrt{ 1 -  \frac{2(m_W^2 + m_{H^{\pm}}^2)}{s}
                        +  \frac{(m_W^2 - m_{H^{\pm}}^2)^2}{s^2}}, \\ 
  \cos \Theta &=& \frac{2 t  + s - m_{H^{\pm}}^2 - m_W^2}{s \beta_{HW}}, 
\end{eqnarray}
where $s$ and $t$ are the Mandelstam variables 
( $s = (k + \overline{k})^2 = (p + \overline{p})^2$, 
  $t = (k - p)^2 = (\overline{k} - \overline{p})^2$). 
The total cross section is calculated according to the formula 
\begin{eqnarray}
  \sigma (s) = \frac{1}{16 \pi} \frac{1}{s^2} \int_{t_{\rm min}}^{t_{\rm max}}
     \frac{1}{2}   \sum_{\tau} \sum_{\overline{\lambda}}  
     \left| {\cal M}
     (k,\overline{k},\tau; p, \overline{p}, \overline{\lambda}) \right|^2
      d t ,  
\end{eqnarray}
where $t_{\rm max}$ and $t_{\rm min}$ are defined by  
\begin{eqnarray}
  t_{\rm max} &=& \frac{1}{2}(m_{H^\pm}^2 + m_W^2 - s + s \beta_{HW}), \\  
  t_{\rm min} &=& \frac{1}{2}(m_{H^\pm}^2 + m_W^2 - s - s \beta_{HW}).   
\end{eqnarray}
Our formalism here is consistent with 
that for $e^-e^+ \to \chi^- W^+$ ($\chi^-$: the charged Goldstone boson) 
in Ref.~\cite{brs} in the limit $m_{H^\pm}^2 \to m_\chi^2$  
and also with that for $e^-e^+ \to H^0 \gamma$ in Ref~\cite{Hgamma}.

In calculation, the form factors $F_{i,\tau}(s,t)$ may be decomposed 
according to each type of Feynman diagrams (Fig. 1) as     
\begin{eqnarray}
  F_{i,\tau}(s,t) = F_{i,\tau}^{\gamma}(s) + F_{i,\tau}^{Z}(s) + 
               F_{i,\tau}^{t}(t) + F_{i,\tau}^{Box}(s,t) + 
               \delta F_{i,\tau}(s,t),  \label{fdec}
\end{eqnarray}
where $F_{i,\tau}^{V}$ ($V = \gamma$ and $Z$) are the contribution 
from the one-loop induced $HWV$ vertices (Fig. 1(a)). 
These $HWV$ vertices are defined as 
$i g m_W V_{\mu\nu}^{HWV}$ (Fig.~2), in which 
$V_{\mu\nu}$ may be expressed by\cite{mp,kanemu}
\begin{eqnarray}
V_{\mu\nu}^{HWV}(\mg,p_W^2,p_V^2) 
&=& F^{HWV}(\mg,p_W^2,p_V^2) \gmn    
+ G^{HWV}(\mg,p_W^2,p_V^2) \fr{{p_V}_{\mu} {p_W}_{\nu}}{\mw} \nn\\
&&  + 
i H^{HWV}(\mg,p_W^2,p_V^2)
 \fr{p_V^{\rho}p_W^{\sigma}}{\mw} \epsilon_{\mu\nu\rho\sigma},  
\end{eqnarray}
where $p_H$ is the incoming momentum of $H^-$, and 
$p_V$ ($V = Z$ or $\gamma$) and $p_W$ are outgoing momenta 
of $V$ and $W$ bosons, respectively. The form factor 
$F_{i,\tau}^{V}(s)$ are then expressed by 
\begin{eqnarray}
  F_{1,\tau}^{V}(s) &=& g m_W C_{V} \frac{1}{s - m_V^2}
        F^{HWV} (m_W^2,s,m_{H^\pm}^2), \\
  F_{2,\tau}^{V}(s) &=& g m_W C_{V} \frac{1}{s - m_V^2}
        \frac{1}{2} G^{HWV} (m_W^2,s,m_{H^\pm}^2), \\
  F_{3,\tau}^{V}(s) &=& g m_W C_{V} \frac{1}{s - m_V^2}
        \frac{-1}{2} H^{HWV} (m_W^2,s,m_{H^\pm}^2) , 
\end{eqnarray}
where $m_V$ is mass of the neutral gauge bosons ($m_Z$ and $m_\gamma (= 0)$),  
and $C_V$ are defined by 
$C_\gamma = e Q_e$ and $C_Z = g_Z (T_e^3 - s_W^2 Q_e)$ 
$(e = g s_W = g_Z s_W c_W)$, where $Q_e = -1$, and $T_e^3 = -1/2$ ($0$) 
for the electron with the helicity $\tau = -1$ ($+1$). 
The explicit formulas of the $F^{HWV}$, $G^{HWV}$, $H^{HWV}$  
are given in Appendix A.~1. 
The $F_{i,\tau}^{t}(s,t)$ is the contribution 
of the $t$ channel diagram with the one-loop 
$H^-W^+$ mixing diagrams (Fig.~1(b)) 
and the box diagram contributions are expressed 
by $F_{i,\tau}^{Box}$ (Fig.~1(c)). 
We also show the explicit results for 
$F_{i,\tau}^{t}$ and $F_{i,\tau}^{Box}$ in Appendix A.~3 and A.~4, 
respectively. 
Each one-loop-diagram contribution to $F_1(s,t)$ except for 
$F_{1,\tau}^{Box}$ includes the ultraviolet divergence. 
After summing up the contributions $F_{i,t}^V$, $F_{i,\tau}^t$, 
and $F_{i,\tau}^{box}$, the divergence is canceled 
out because of no tree-level contribution. 

Although the amplitude is finite already, by making the renormalization 
for the $WH$ and $wH$ two point functions the finite counterterm, 
$\delta F_{i, \tau}$, is introduced to this 
process\cite{reno,mix-cap}\footnote{See also {\bf Note Added.}}. 
By rewriting the fields $w^\pm$ and $H^\pm$ with shifting 
$\beta \rightarrow \beta - \delta \beta$ by 
\begin{eqnarray}
   \left(\begin{array}{c}
         w^\pm \\ H^\pm 
   \end{array}\right)
&\rightarrow& \left(\begin{array}{cc}
         Z_{w^\pm}^{\fr{1}{2}} &  Z_{wH}^{\fr{1}{2}}\\ 
         Z_{Hw}^{\fr{1}{2}}    & Z_{H^\pm}^{\fr{1}{2}}
   \end{array}\right)
\left(\begin{array}{cc}
         1 &  - \delta \beta\\ 
         \delta \beta    & 1
   \end{array}\right)
   \left(\begin{array}{c}
         w^\pm \\ H^\pm 
   \end{array}\right) \nn\\ &\equiv&
\left(\begin{array}{cc}
         1+\frac{1}{2} Z_{w^\pm}^{(1)} &  a^{(1)}_{wH}\\ 
         a^{(1)}_{Hw}    & 1+\frac{1}{2} Z_{H^\pm}^{(1)}
   \end{array}\right)
   \left(\begin{array}{c}
         w^\pm \\ H^\pm 
   \end{array}\right),  \label{shift}
\end{eqnarray}
the relevant counterterms are extracted from the kinematic terms of the
Higgs sector as 
\begin{eqnarray}
\!\!\!\!\!\!\!  
{\cal L}^{count.}\!\! &=& \!\! i \, a_{wH}^{(1)}  \frac{g v}{2} 
  W_\mu^- \partial^\mu H^+
- a_{wH}^{(1)}  \frac{g^2 v}{2} \frac{s_W^2}{c_W} 
  W_\mu Z^\mu H^+ 
+ a_{wH}^{(1)}  \frac{g^2 v}{2} s_W 
  W_\mu \gamma^\mu H^+ + {\rm h.c.}. \label{count}
\end{eqnarray}
For the $WH$ mixing we take the renormalization condition 
\begin{eqnarray}
  {\rm Re} \left( \Pi_{WH}^{\rm reno.}(m_{H^{\pm}}^2) \right) = 
  {\rm Re} \left( \Pi_{WH} (m_{H^\pm}^2) \right) + 
                  \Pi_{WH}^{\rm count.} = 0,   
\end{eqnarray}
where $\Pi_{WH}(p^2)$ is given in Eq.~\eq{pi-WH} in Appendix. we then obtain 
\begin{eqnarray}
       a_{wH}^{(1)} = \frac{1}{m_W^{}}   
          {\rm Re} \left( \Pi_{WH} (m_{H^\pm}^2) \right),  \label{awH}
\end{eqnarray}
so that the counterterms for not only the $WH$ mixing but also 
the $HWV$ vertices are obtained by using Eq.~\eq{count}.    
Next, \eq{shift} also produces the $wH$ mixing 
($w$: the charged Goldstone boson). We fix the counterterm 
so as to satisfy the renormalization condition\cite{mix-cap}   
\begin{eqnarray}
  {\rm Re} \left( \Pi_{Hw}^{\rm reno.}(m_{H^{\pm}}^2) \right)= 
  {\rm Re} \left( \Pi_{Hw} (m_{H^\pm}^2) \right) + 
                  \Pi_{Hw}^{\rm count.}
 = 0. 
\end{eqnarray}
The finite counterterms for the form factors,   
$\delta F_{i,\tau}$ in Eq.~\eq{fdec}, are then obtained 
as we show in Appendix A.5.

\section{Non-decoupling mass effects}

\hspace*{18pt}
Here we present some analytic discussion on the amplitudes to 
find cases in which the cross section becomes large for a given 
$\sqrt{s}$ in the non-SUSY 2HDM.  

Let us consider the quark-loop contributions to the amplitudes first. 
They do not decouple in the heavy quark 
limit because the decoupling theorem\cite{dec} does not work 
for the Yukawa interactions in which the couplings are proportional 
to the squared masses. Hence larger one-loop effects take place 
for heavier quark masses\footnote{We here call them as 
the non-decoupling effects.}. 
In the helicity amplitude with a longitudinally polarized $W$ boson,   
powerlike top- or bottom-quark mass contributions appear by the factor of 
$m_t^2 \cot\beta$ or $m_b^2\tan\beta$ in Model II.  
The linear appearance of $\cot\beta$ or $\tan\beta$ in each factor 
comes from the fact that one $tb H^\pm$ Yukawa coupling is included 
in each $t$-$b$ loop diagram\footnote{The $tbH^-$ coupling gives 
$m_t \cot\be$ and $m_b \tan\be$, and the other 
$m_t$ and $m_b$ comes from the $tbW_L^+$ coupling ($W_L$ represent 
the longitudinal $W$ boson).  By the chirality argument other combinations 
such as $m_t m_b \cot \beta$ and $m_t m_b \cot\beta$ disappear.}.
Each factor becomes large for small $\tan \beta$ 
($\ll \sqrt{m_t/m_b}$) or for large $\tan\be$ 
($\gg \sqrt{m_t/m_b}$), respectively.
In our analysis, we take into account theoretical lower and upper 
bounds of $\tan\be$ putting a criterion for the upper limit of 
the top-Yukawa coupling $y_t$ ($\propto m_t/\sin\be$) and 
the bottom-Yukawa coupling $y_b$ ($\propto m_b/\cos\be$) 
by the requirement for validity of perturbation theory.   
Under the same criterion for both top- and bottom-Yukawa coupling constants, 
the factor $m_t^2 \cot \beta$ at the lowest $\tan\be$ value is by 
$m_t/m_b$ greater than the factor $m_b^2\tan\beta$ at the highest 
$\tan\beta$ value.    
Therefore the helicity amplitude becomes large especially for small $\tan\be$ 
($\ll \sqrt{m_t/m_b}$) by the $t$-$b$ loop contributions\footnote{
Similar top-bottom quark effects are observed in the cross
section of $e^+e^- \rightarrow A^0 V$ ($V= \gamma, Z^0$)\cite{eeav}.}. 
In  Model I, $\tan \beta$ is just replaced by $\cot \beta$ in the 
coefficient above, hence this change does not affect on above discussion. 
Therefore in both Model I and II, we 
expect to have sizable cross sections for small $\tan \beta$ values.

Next we discuss the Higgs-loop contributions. The non-decoupling effects 
of the heavy Higgs bosons appear only when the Higgs sector has a special 
property:  
the Higgs mass squared are expressed like $\sim \lambda_i v^2$, 
where $\lambda_i$ is a combination of the Higgs self-coupling constants. 
This corresponds to $M \ll v$ in our notation\cite{kanemu,kko}, where $M$ 
is the scale of the soft breaking of the discrete symmetry.   
In this case, similarly to the Yukawa interaction, the terms of 
${\cal O}(m_{H_i^0}^2/v^2)$ appear in the helicity amplitude with a 
longitudinally polarized $W$ boson, 
where $H_i^0$ represent heavy neutral Higgs bosons in the loop. 
Therefore, in the non-SUSY 2HDM with the small soft-breaking mass $M$,
these mass effects of heavy Higgs bosons may enhance the amplitude 
in addition to the $t$-$b$ loop effects. 
Clearly, this situation is quite different from the MSSM like Higgs sector, 
where large masses of the extra Higgs bosons are possible only by taking 
large $M$ ($\gg \lambda_i v^2 = {\cal O}(g^2 v^2)$) 
\footnote{In the MSSM, $m_A$ corresponds to $M$.}. 

In order to see the leading non-decoupling effects  
(the quadratic-mass terms in the large mass limit for particles in the loop) 
analytically, let us consider the amplitude with a longitudinally 
polarized $W$ boson in one limiting case. 
They are extracted from the full expression of the 
amplitude by taking masses of $h^0$, $H^0$ and $A^0$ much larger than 
$m_W$ and $m_{H^\pm}$ with setting $M=0$  \footnote{
This expression is for the $\delta F_{i,\tau}=0$ case.};   
\begin{eqnarray} 
&&
\!\!\!\!\!\!\!\!\!\!\!\!
\!\!\!\!\!\!\!\!\!\!\!\!
 {\cal M} (k,\overline{k},\tau; p, \overline{p}, \overline{\lambda}=0)
   \nn\\
&&
\!\!\!\!\!\!\!\!\!\!\!\!
\!\!\!\!\!\!\!\!\!\!\!\!
= \sin \Theta \frac{g^2}{c_W^2} \frac{T_e^3}{16 \pi^2 v^2}
\left[ 
\frac{3}{2} \left\{ \frac{\mH \ma}{\mH - \ma} \ln \frac{\mH}{\ma} 
                  - \frac{\mh \ma}{\mh - \ma} \ln \frac{\mh}{\ma} 
            \right\}  J(\al, \be) \right. \nn\\
&&       
  - \left\{ \frac{c_{2W}}{2} \mH  
                 + \frac{3}{4} \frac{\mH \ma}{\mH - \ma} \ln \frac{\mH}{\ma} 
                  \right\}    K(\al, \beta)  \nn\\
&& \left.
     -  \left\{   \frac{c_{2W}}{2} \mh  
                + \frac{3}{4} \frac{\mh \ma}{\mh - \ma} \ln \frac{\mh}{\ma} 
        \right\}    L(\al, \beta) 
-  \frac{N_c}{2} \mt \cot \be
\right]  \nn \\
\!\!\!\!\!\!\!\!\!\!\!\!
\!\!\!\!\!\!\!\!\!\!\!\!
&&
\!\!\!\!\!\!\!\!\!\!\!\!
\!\!\!\!\!\!\!\!\!\!\!\!
+ \sin \Theta  \frac{g^2 s_W^2 Q_e}{16 \pi^2 v^2}
\left[  \frac{3}{2} \left\{ \frac{\mH \ma}{\mH - \ma} \ln \frac{\mH}{\ma} 
                     - \frac{\mh \ma}{\mh - \ma} \ln \frac{\mh}{\ma} 
               \right\}  J(\al, \be)  \right. \nn\\
\!\!\!\!\!\!\!\!\!\!\!\!
\!\!\!\!\!\!\!\!\!\!\!\!
&&
- \left\{  \frac{1}{2 c_W^2} \mH  
              - \frac{3}{4} \frac{\mH \ma}{\mH - \ma} \ln \frac{\mH}{\ma} 
                 \right\}    K(\al, \beta)
 \nn\\
\!\!\!\!\!\!\!\!\!\!\!\!
\!\!\!\!\!\!\!\!\!\!\!\!
&& \left.
   -  \left\{    \frac{1}{2 c_W^2} \mh  
              - \frac{3}{4} \frac{\mh \ma}{\mh - \ma} \ln \frac{\mh}{\ma} 
                  \right\}    L(\al, \beta)  +
   \frac{N_c}{2 c_W^2} \mt \cot \be
\right] \, + \,  {\cal O} \left( \frac{s}{m_{H^0_i}^2} \right),
 \label{quad} 
\end{eqnarray}
where $H_i^0$ represents $h^0$ $H^0$ and $A^0$, and  
\begin{eqnarray}
J(\al,\be) &=& \snab \csab,            \label{fac2}                \\
K(\al,\be) &=& \sin^2 \al \ctb - \cos^2 \al \tnb,   \label{fac3}   \\
L(\al,\be) &=& \cos^2 \al \ctb - \sin^2 \al \tnb.   \label{fac4}
\end{eqnarray}
From the expression \eq{quad}, we expect that the amplitude can become 
large by the non-decoupling effects of the heavy Higgs bosons as well as 
those of the top-quark. The Higgs effects grow for the 
large or small $\tan\beta$: see \eq{fac2}-\eq{fac4}. 

The non-SUSY 2HDM receives rather strong theoretical constraints.  
First from the requirement for validity of perturbation theory, 
all the Higgs-self coupling and Yukawa coupling constants should not 
be so large\cite{uni,tri,tanb}. 
We here set a rather conservative criterion corresponding to Ref.~\cite{kko}: 
that is, for the Yukawa couplings  
\begin{eqnarray}
   y_b^2, \; y_t^2 &<& 4 \pi,    \label{uniyuk}
\end{eqnarray}
and for the Higgs self-coupling constants  
\begin{eqnarray}
   |\lambda_1|, \; |\lambda_2|, \; |\lambda_3|, \;   
   \frac{1}{4} |\lambda_4 \pm \lambda_5|  &<& 4 \pi .   
  \label{unilam}
\end{eqnarray} 
These conditions give constraints on the combinations 
among masses, mixing angles and the soft-breaking mass.
For example, from the condition for $\lambda_1$, we obtain 
by using \eq{qu1}   
\begin{eqnarray}
   (\mH - M^2) \tan^2 \be \lsim 8 \pi v^2,  \label{unitarity}
\end{eqnarray}
for the case of $\al = \be - \pi/2$ and $m_{H^0}^2 \gg m_{h^0}^2$. 
This means that it is difficult 
to take large $m_{H^0}$ and large $\tan \beta$ simultaneously with 
$M^2 \sim 0$. We include all these constraints in our numerical analysis. 

Finally the 2HDM is constrained from the precision experimental data\cite{gr}, 
especially those for the $\rho$ parameter: 
the additional contribution of the 2HDM Higgs sector should be small. 
We here employ the same condition as in Ref.~\cite{kko};  
$\Delta \rho_{\rm 2HDM} = - 0.0020 - 0.00049 \frac{m_t - 175 {\rm GeV}} 
 {5 {\rm GeV} } \pm 0.0027$. 
In order to satisfy this there are mainly two kinds of 
possibility for the parameter choice. 
A)  The Higgs sector is custodial $SU(2)_V$ symmetric ($\mg \sim \ma$). 
B)  The Higgs sector is not custodial $SU(2)_V$ symmetric  
    but there are some relations among parameters 
    to keep a small $\Delta \rho_{\rm 2HDM}$:  
    $\mg \sim \mH$ or $\mg \sim \mh$ with $\al \sim \be - \pi/2$ 
    or $\al \sim \be$, respectively\cite{hhg}. 
Also, recent study for the $b \rightarrow s \gamma$ results\cite{bsg_ex} 
gives the constraint on the charged Higgs boson mass 
($m_{H^\pm} \gsim 160$ GeV)\cite{bsg_mg}.

By taking into account all the theoretical and experimental constraints above, 
the best choice for the maximal Higgs contributions to the cross section 
is to take the case B) and then to choose $m_{A^0}$ and $\tan \beta$ 
as large as possible under the conditions \eq{uniyuk} and \eq{unilam}.

\section{Numerical Evaluation}

\hspace*{18pt}
We here show our numerical results. According to the above analytic 
discussion, the 7 free parameters of the Higgs sector 
in the non-SUSY 2HDM ($\mh$, $\mH$, $\mg$, $\ma$, $\al$, $\be$ and $M$) 
are chosen in the following way. 
To obtain larger Higgs contributions, we take the choice B) 
in the last section. Since $m_{h^0} < m_{H^0}$, it is better to set 
$\alpha = \beta - \pi/2$ (or $\al =0$) for larger cross section for 
$\tan \beta > 1$ ($K(\al,\be) > 1$) (See \eq{fac3}). 
If we choose $\alpha = \beta$  (or $\al = \pi/2$), then such enhancement 
takes place for small $\tan \beta$ ($L(\al,\be) \sim 1$). 
Any other choice of $\al$ leads to smaller cross sections. 
As for the quark loops, although we here adopt Model II for the 
Yukawa couplings in actual calculation in the 2HDM, it is clear that 
there is no difference between Model I and II for the cross section 
except for the large $\tan\be$ regime.
If we assume the MSSM Higgs sector, there are two free parameters  
$m_{H^\pm}$ and $\tan \be$, and all the other parameters 
are related to these two parameters\cite{hhg}. 
As for the quark masses we here fix them as $m_t =175$ GeV and $m_b=5$ GeV.

To begin with, we show the total cross section for $m_{H^\pm} = 200$ GeV 
at $\sqrt{s} = 500$ GeV as a function of $\tan\be$ (Fig.~3). 
The region of $\tan\be$ is $0.28 < \tan\be < 123$ taking into account 
the condition \eq{uniyuk}\footnote{
As for the constraint for $\tan\be$ in the MSSM, see 
Refs.~\cite{hhw,DELPI,susytb}.
}, while we switch off the condition \eq{unilam} 
in Fig.~3 (and in Fig.~4) just to concentrate on showing the behavior of 
the non-decoupling effects more clearly. 
The results in which both the conditions \eq{uniyuk} and \eq{unilam}  
are included will be shown soon later in Figs.~5 and 6.   
In Fig.~3, the real curves represent the total cross sections 
in the non-SUSY 2HDM for each value of $m_{A^0}$.  The other parameters 
are taken as $m_{h^0} = 120$ GeV, $m_{H^0} = 210$ GeV, 
$\alpha = \beta - \pi/2$ and $M=0$.
The dotted curve represents the cross section in the MSSM 
with super-heavy super-partner particles. 
For small $\tan\beta$ ($\ll \sqrt{m_t/m_b}$), as we discussed in the 
last section, the cross section is enhanced by the $t$-$b$ loop 
contributions both in the MSSM and in the non-SUSY 2HDM.
On the other hand, for large $\tan \beta$ ($\gg \sqrt{m_t/m_b}$), 
the MSSM cross section reduces rapidly, while the Higgs non-decoupling 
effects enlarge the non-SUSY 2HDM cross section. For larger $m_A$, 
larger cross sections are observed.       
Our result in the MSSM here is consistent with that in Ref.~\cite{zhu}. 

Fig.~4 shows the $\sqrt{s}$ dependence of the total cross section 
in the non-SUSY 2HDM at $m_{H^\pm} = 200$ GeV for various $\tan \be$, 
other parameters are chosen as $m_{h^0} = 120$ GeV, 
$m_{H^0} = 210$ GeV, $m_{A^0} = 1200$ GeV and 
$\alpha = \beta - \pi/2$ and $M=0$. 
The condition \eq{unilam} is switched off in this figure too.

The enhancement of the cross section essentially depends on the size of 
the $H^\pm tb$ and $H^\pm H^\mp H^0$ coupling constants. By taking these 
couplings as large as possible under the conditions \eq{uniyuk} and 
\eq{unilam} and also under the experimental constraints mentioned before, 
we obtain upper bounds of the cross section in the non-SUSY 2HDM 
for each value of $m_{H^\pm}$ and $\tan\be$. 
The situation is described in Fig.~5. 
The dotted curve represents the cross section with $M=0$ at 
$\sqrt{s} = 500$ GeV for $m_{H^\pm}=200$ GeV at 
$\al = \be - \pi/2$, 
and all the other free parameters in the Higgs sector are chosen 
in order to obtain maximum Higgs non-decoupling effects under all 
the conditions\footnote{The other choice of $\al$ leads to less Higgs 
effects for $\tan\be > 1$ in this case.}. 
For $\tan\be \gsim 5.9$, the condition \eq{unitarity} obtained from 
\eq{unilam} cannot be satisfied any more if we keep $M=0$: 
larger value of $\tan \beta$ is allowed only by introducing nonzero 
soft-breaking mass $M$. This leads to the smaller cross section 
because the non-decoupling property of the Higgs sector is weakened 
by non-zero $M$: see the discussion in Sec.~4. 
Therefore the upper bounds are obtained as the solid curve. 
The cross section rapidly reduces for $\tan\be \gsim 5.9$. 
Although the quark-loop contributions (the bottom mass effects) 
enhance the cross section for $\tan\be \gsim 40$, 
the magnitude is still much less than that for small $\tan\be$. 

In Fig.~6 we show such general bounds of the cross section as a function 
of $\tan \be$ at $\sqrt{s} = 500$ GeV for 
$m_{H^\pm} = 160, 200, 240, 280, 320$ and $360$ GeV.    
All the other free parameters are chosen as the same way in Fig.~5. 
Each peak of the cross section in the moderate $\tan \be$ value is the point 
where the largest Higgs non-decoupling effects with $M = 0$ appear.

\section{Discussion and Conclusion}

\hspace*{18pt}
We have discussed the $H^\pm$ production process {\it via} 
$e^+e^- \rightarrow H^\pm W^\mp$ in the non-SUSY 2HDM as well as in the MSSM.

In the non-SUSY 2HDM, the large cross section is possible for small 
$\tan\be$ by the $t$-$b$ loop contributions (quadratic top-mass effects). 
At $\tan\be=0.3$, for $m_{H^\pm} = 200$ GeV, the cross section can be as 
large as 8 fb at $\sqrt{s}=500$ GeV and maximally it reaches to over 40 fb 
at $\sqrt{s}\sim390$ GeV. 
For larger $\tan\be$, these top-mass effects decrease until 
$\tan\be\sim m_t/m_b = 35$. In Model II, the quadratic bottom-mass 
effects enhance the cross section for $\tan\be \gsim m_t/m_b$, but the 
magnitude is not so large: 
at $\sqrt{s}=500$ GeV it is at most a few times $10^{-2}$ fb even for 
$\tan\be\sim 100$. If Model I is assumed, this small enhancement for 
$\tan\be > m_t/m_b$ disappears, but all the results for smaller $\tan\be$ 
are almost same as those in Model II.
In addition to the quark-loop effects, the Higgs non-decoupling 
effects contribute to the cross section by a few times 0.1 fb  
for moderate values of $\tan \beta$. 
Such Higgs effects are strongly bounded for larger $\tan \be$ 
($\gsim  \sqrt{m_t/m_b}$) 
by the requirement for validity of perturbation theory. 

In the MSSM with heavy super-partner particles, the Higgs-loop 
effects are very small and only the $t$-$b$ loops 
contribute to the cross section. 
For $m_{H^\pm} = 200$ GeV, the cross section at $\tan\be=2$ amounts to 
a few times 0.1 fb at $\sqrt{s}=500$ GeV, and maximally it reaches to over 
1 fb at $\sqrt{s} \sim 390$ GeV. 
The cross section rapidly reduces for larger $\tan\beta$. 
We here have not discussed the one-loop contributions of the super-partner 
particles in the MSSM explicitly, which will be discussed in our future paper. 

We give some comments on our analysis. 
First, our results have been tested in the high-energy limit 
by using the equivalence theorem\cite{et} at one-loop level\cite{et2}.
We evaluated $e^-e^+ \to H^- w^+$ ($w^+$: the charged Goldstone boson) 
and confirmed that the cross section was coincident with our prediction for 
the $H^-W^+_L$ production in the high-energy limit.    
Second, although the process is one-loop induced and so the 
ultraviolet divergences have canceled among the diagrams, 
we have include the finite renormalization effects of the $WH$ mixing 
and the $w H$ mixing by putting the renormalization conditions on the 
mass shell of $H^\pm$. The effects have turned out to give a few \% 
(at most about 5\%) of corrections to the one-loop-induced cross sections 
in which the finite renormalization effects ($\delta F_{i,\tau}$) 
are not included.    

Finally we comment on detectability of the signal events for the
case of $m_{H^\pm} > m_t + m_b$. The $H^\pm$ decays into a $tb$
pair and the signal process is 
$e^+e^- \rightarrow H^\pm W^\mp \rightarrow t \bar{b} W^- + \bar{t} b W^+$. 
The main background process may be 
$e^+e^- \rightarrow t \bar{t} \rightarrow t \bar{b} W^- + \bar{t} b W^+$. 
The cross section of $e^+e^- \rightarrow t \bar{t}$ amounts to about 
0.57 pb for $\sqrt{s} = 500$ GeV: 
the signal/background ratio is at most around 1 \%.  
It may be, however, expected that the signal can be comfortably seen 
if the signal cross section is 10fb, by attaining a background 
reduction in Ref.~\cite{oda-pr} by the following method: 
1) cut around reconstructed $b W$ masses which can come from $b W$ decay 
   at $175$GeV, 
2) find a peak in reconstructed $m_{H^\pm}$ and 
3) confirm the presence of $H^\pm$ according to the method in 
Ref.~\cite{oda-top}. 
For smaller signal cross sections of the order of 0.1fb,  details of 
the background analysis are needed to see the detectability.

\subsection*{Note added:}

\hspace*{18pt}
After this work was finished, another paper (Ref.~\cite{achm}) appeared  
in which the same subject was studied. 

\section*{\it Acknowledgments}

\hspace*{18pt}
The author would like to thank W.~Hollik for useful discussion, 
K.~Odagiri for valuable discussion about the backgrounds and the 
detectability at LC, Y. Okada for useful information about 
the constraint on $\tan\beta$. This work was supported, in part,  
by the Alexander von Humboldt Foundation.
\newpage

\appendix

\section{Analytic results}

\hspace*{18pt}
In the formulas here, we use the integral functions introduced by 
Passarino and Vertman\cite{pave}. The notation for the tensor coefficients 
here is based on Ref.~\cite{brs}. We here write $A(m_f)$ as $A[f]$,  
$B_{ij}(p_H^2; m_{f_1}, m_{f_2})$ as $B_{ij}[f_1, f_2]$, 
$C_{ij}(p_H^2,\pwsq,p_V^2; m_{f_1}, m_{f_2}, m_{f_3})$ as 
$C_{ij}[f_1, f_2, f_3]$, where  
$f_i$ are the fields with mass $m_{f_i}$. For the quark diagrams, 
we define abbreviation 
$C_{ij}(tbb) = C_{ij}(p_H^2,\pwsq,p_V^2; m_t, m_b, m_b)$ and 
$C_{ij}(ttb) = C_{ij}(p_H^2,p_V^2,\pwsq; m_t, m_t, m_b)$.  
The exression is in the 't Hooft-Feynman gauge. 
Also $J(\al,\be)$, $K(\al,\be)$ and $L(\al,\be)$ 
in Eqs.~\eq{fac2} - \eq{fac4} are written as 
$J_{\al\be}$, $K_{\al\be}$ and $L_{\al\be}$, and we also write 
\begin{eqnarray}
  \tilde{K}_{\al\be} &=& \left\{  K_{\al\be} (\mH - M^2) 
                              -  J_{\al\be} (2 \mg - \mH)\right\},  \\
  \tilde{L}_{\al\be} &=& \left\{  L_{\al\be} (\mH - M^2) 
                              +  J_{\al\be} (2 \mg - \mh)\right\}, 
\end{eqnarray}
respectively, for brevity.
The momentum squared of the $H^+$ is set on mass-shell, $p_H^2 = \mg$.

\subsection{Form factors of the $H^+ W^- V^0$ ($V^0 = Z^0$, $\gamma$) Vertices}

\hspace*{18pt}
We write each contribution to the unrenormalized $H^\pm W^\mp V^0$ 
form factors $X^{HWV}$, ($X = F, G$ and $H$) as 
$X^{HWV} = X^{{HWV}(a)} + X^{{HWV}(b)} +X^{{HWV}(c)}$ corresponding 
to Figs.~7(a), 7(b) and 7(c).  
$X^{{HWV}(a)}$ is the contribution of triangle-type diagrams (Fig.~7(a)), 
$X^{{HWV}(b)}$ represents that from the two-point function correction 
shown in Fig.~7(b), and $X^{{HWV}(c)}$ 
is tadpole contribution as well as some two-point function 
corrections written only by the $A$ function (Fig.~7(c)).

\subsubsection{The $H^+ W^- Z^0$ vertex}

\hspace*{18pt}
The contribution of triangle-type diagrams to $F^{HWZ}$ is calculated as 
{\small 
\begin{eqnarray}
&& \!\!\!\!\!\!\!\!\!\!\!\! 
  F^{HWZ (a)}(\mg,p_W^2,p_Z^2) = \fr{2}{16 \pi^2 v^2 c_W} \nn \\ 
&& \!\!\!\!\!\!\!\!\!\!\!\! 
   \times \lt[  - \tilde{K}_{\al\be}   
              \left\{ C_{24}[H^\pm A^0H^0] 
               - c_{2W} C_{24}[H^0H^\pm H^\pm]  \right\}
     -  \tilde{L}_{\al\be}
\lt\{C_{24}[H^\pm A^0h^0] - c_{2W} C_{24}[h^0H^\pm H^\pm] \rt\} 
\rt.  \nn \\
&& \!\!\!\!\!\!\!\!\!\!\!\! 
     + J_{\al\be}\lt\{ 
  (\mg - \mH) C_{24} \lt( [w^\pm z^0 H^0] - c_{2W} [H^0 w^\pm w^\pm] \rt) 
     -  (\mg - \ma) C_{24}  [w^\pm H^0 A^0]   \rt.\nn \\
&& \!\!\!\!\!\!\!\!\!\!\!\!  
     -  \mw C_{24}   [W^\pm H^0 A^0] -  \fr{c_{2W}}{c_W} 
       m_W^2 C_{24}  [H^\pm H^0 Z^0] 
     -  \mw
     \lt( 4 (p_W^2 + p_W \cdot p_Z ) C_0 
     + 2 (2 p_W + p_Z) \cdot  \rt. \nn\\ 
&& \!\!\!\!\!\!\!\!\!\!\!\!  
 \lt. (p_W C_{11} + p_Z C_{12})+ p_W \cdot p_Z C_{23} + (D-1) C_{24} \rt)
      [W^\pm Z^0 H^0]  
     +  c^2_W \mw 
     \lt( (p_Z^2 - p_W^2 ) C_0 
      \rt. \nn\\
&& \!\!\!\!\!\!\!\!\!\!\!\!  \lt. 
- 2 p_Z \cdot (p_W C_{11} + p_Z C_{12}) 
+ p_W \cdot p_Z C_{23} + (D-1) C_{24} \rt)
      [H^0 W^\pm W^\pm] 
     -  \mz (\mg - \mH) s^2_W C_0[w^\pm Z^0 H^0] \nn \\
&& \!\!\!\!\!\!\!\!\!\!\!\! 
   \lt.\lt.  -  \mw (\mg - \mH) s^2_W C_0[H^0 W^\pm w^\pm] 
     +  \mw s^2_W 
        C_{24}[H^0 w^\pm W^\pm]  - (H^0 \rightarrow h^0) \rt\} \rt] \nn\\
&&\!\!\!\!\!\!\!\!\!\!\!\!
+ \fr{4 N_c}{16 \pi^2 v^2 c_W}
\lt[ \mb \tan \be 
\lt\{ (- s_W^2 Q_b) 
\lt( \pw \cdot (\pw + \pz) C_{11} + \pz \cdot (\pw + \pz) C_{12} 
    + \pwsq C_{21} + \pzsq C_{22}  
 \rt.\rt.\rt. \nn\\
&& \!\!\!\!\!\!\!\!\!\!\!\!
\lt. + 2 \pw \cdot \pz C_{23} + D C_{24} \rt)(tbb)   -  (T_b- s_W^2 Q_b)
 \lt( \pwsq  C_{11} + \pz \cdot \pz C_{12} 
+ \pwsq C_{21} + \pzsq C_{22} + 2 \pw \cdot \pz C_{23}\rt. \nn\\
&& \!\!\!\!\!\!\!\!\!\!\!\!
\lt. + (D-2) C_{24} \rt)(tbb) - (T_t- s_W^2 Q_t)
 \lt( \pzsq  C_{11} + \pz \cdot \pz C_{12} 
+ \pzsq C_{21} + \pwsq C_{22} + 2 \pw \cdot \pz C_{23} \rt.  \nn\\
&& \!\!\!\!\!\!\!\!\!\!\!\! 
\lt.\lt.+ (D-2) C_{24} \rt)(ttb) +  (- s_W^2 Q_t) \mt C_0(ttb) \rt\}
+ \mt \cot \be \left\{ - (T_b- s_W^2 Q_b)
 \lt(  (\pwsq + \pw \cdot \pz )   C_{0} 
\rt.\rt. \nn\\
&& \!\!\!\!\!\!\!\!\!\!\!\! 
\lt.      +  (2 \pwsq + \pw \cdot \pz ) C_{11} 
     +  (\pzsq + 2 \pw \cdot \pz ) C_{12} + \pwsq C_{21} 
+ \pzsq C_{22} + 2 \pw \cdot \pz C_{23} + (D-2) C_{24} 
        \rt)(tbb)  \nn\\
&& \!\!\!\!\!\!\!\!\!\!\!\!  
+   (- s_W^2 Q_b) \mb C_0(tbb) 
+  (- s_W^2 Q_t)  
     \lt( \pz \cdot (\pw + \pz) C_{11} + \pw \cdot (\pw + \pz) C_{12} 
    + \pzsq C_{21} + \pwsq C_{22} \rt.\nn\\
&& \!\!\!\!\!\!\!\!\!\!\!\! 
\lt. + 2 \pw \cdot \pz C_{23} + D C_{24} 
        \rt)(ttb)  -  (T_t- s_W^2 Q_t)
 \lt( (\pzsq + \pw \cdot \pz ) C_{0} 
     +  (2 \pzsq + \pw \cdot \pz ) C_{11}\rt. \nn\\
&& \lt.\lt. + (\pwsq + 2 \pw \cdot \pz ) C_{12} + 
\pzsq C_{21} + \pwsq C_{22} + 2 \pw \cdot \pz C_{23} + 2 C_{24} \rt)(ttb)   
\rt].
\end{eqnarray}
}
The contribution of the diagrams  expressed in terms of 
the $B_i$ functions is given by 
{\small 
\begin{eqnarray}
&& \!\!\!\!\!\!\!\!\!\!\!\! 
      F^{HWZ (b)}(\mg,p_W^2,p_Z^2) = \fr{2}{16 \pi^2 v^2 c_W}
\lt[  \fr{1}{2}  \tilde{K}_{\al\be}   
\lt\{ s^2_W 
          B_0 [H^0 H^\pm]  
+
\fr{p_Z^2 - p_W^2}{\mg - \mw} c^2_W
        (B_0 + 2 B_1) [H^0 H^\pm]  \rt. \rt. \nn \\
&& \!\!\!\!\!\!\!\!\!\!\!\! 
 \lt. +  \fr{\mH - \mg}{\mg - \mw} s^2_W
        B_0 [H^0 H^\pm]  \rt\}
     + \fr{1}{2}  \tilde{L}_{\al\be} \lt\{  s^2_W 
          B_0 [h^0 H^\pm] 
+ \fr{p_Z^2-p_W^2}{\mg - \mw} c^2_W
        (B_0 + 2 B_1) [h^0 H^\pm] \rt. \nn \\
&& \!\!\!\!\!\!\!\!\!\!\!\! 
\lt.  +
   \fr{\mh - \mg}{\mg - \mw} s^2_W
        B_0 [h^0 H^\pm]  \rt\}  
     + \fr{1}{2} J_{\al\be} \left\{ - (\mg - \mH) s^2_W 
          B_0 [H^0 w^\pm]  
     +   \mw s^2_W  
         B_0 (p_W^2; W^\pm H^0)  
\rt. \nn \\
&& \!\!\!\!\!\!\!\!\!\!\!\! 
     +   \mz s^2_W B_0 
         B_0 (p_Z^2; Z^0 H^0) 
     - \fr{1}{2}  \fr{m_W^2}{\mg - m_W^2} s^2_W
      \lt\{ \mg (B_0 - 2 B_1 + B_{21}) + D B_{22} \rt\}[H^0 W^\pm]   \nn\\
&& \!\!\!\!\!\!\!\!\!\!\!\! 
     + \fr{1}{2}  \mH  
       \fr{\mH - \mg}{\mg - \mw} s^2_W
        B_0 [H^0 w^\pm]   
     +  m_W^2  
       \fr{p_Z^2 - p_W^2}{\mg - \mw} c^2_W
        (B_0 - B_1)  [H^0 W^\pm]  \nn \\
&& \!\!\!\!\!\!\!\!\!\!\!\! 
\lt.\lt.     +    
       \fr{\mH - \mg}{\mg - \mw} (p_Z^2 - p_W^2) c^2_W
        (B_0 + 2 B_1) [H^0 w^\pm]  - (H^0 \rightarrow h^0) \rt\}\rt] \nn\\
&& \!\!\!\!\!\!\!\!\!\!\!\! 
+ \fr{4 N_c}{16 \pi^2 v^2 c_W}
\lt[ \fr{s_W^2}{\mg - m_W^2} 
     \left\{ 
      (\mb \tan \be - \mt \cot \be) 
        \lt( \mg (B_{1} + B_{21}) + D B_{22} \rt) [tb] \rt.\rt. \nn\\
&& \!\!\!\!\!\!\!\!\!\!\!\!  \lt.\lt.
    - \mt \mb (\tan \be - \cot \be) B_0 [tb] \rt\} 
- \fr{c_W^2}{\mg - m_W^2} (\pzsq - \pwsq)
 \lt\{ \mb \tan \be B_1 + \mt \cot \be (B_1 + B_0) \rt\}[tb] \rt] .\nn\\  
\end{eqnarray}
}
The diagrams relevant to the $A$-function is expressed by 
{\small 
\begin{eqnarray}
  && \!\!\!\!\!\!\!\!\!\!\!\! \!\!\!\!\!  \!\!\!\!\! \!\!\!\!\! 
     F^{HWZ (c)}(\mg,p_W^2,p_Z^2) = \fr{1}{16 \pi^2 v^2 c_W} 
  \fr{1}{\mg - \mw} 
\lt[ \;\; s_W^2 \left( 
     \tilde{\Pi}_{Hw}^B  - T_1 \rt) 
-  \lt\{ {s}_W^2 m_W^2 - c^2_W (\pzsq - m_W^2)\rt\} T_2  \rt], \nn\\
\end{eqnarray}
}
where $\tilde{\Pi}_{Hw}^B$,  $T_1$ and $T_2$ are given  
in \eq{tadB}, \eq{tadA} and \eq{two}.

The contribution of the triangle type diagrams to  $G^{HWZ}$ and 
$H^{HWZ}$ are given by 
{\small 
\begin{eqnarray}
  && \!\!\!\!\!\!\!\!\!\!\!\! 
G^{HWZ (a)}(\mg,p_W^2,p_Z^2) = \fr{2 \mw}{16 \pi^2 v^2 c_W} 
 \lt[ -  \tilde{K}_{\al\be}
  (C_{12} + C_{23}) \lt\{    [H^\pm A^0H^0] 
     -  c_{2W} [H^0H^\pm H^\pm] \rt\} 
\rt. \nn \\
&& \!\!\!\!\!\!\!\!\!\!\!\! 
     - \tilde{L}_{\al\be}
    (C_{12} + C_{23}) \lt\{  [H^\pm A^0h^0] - c_{2W} [h^0H^\pm H^\pm] \rt\} 
\nn \\ 
&& \!\!\!\!\!\!\!\!\!\!\!\! 
     + J_{\al\be} \left\{  
(\mg - \mH)(C_{12} + C_{23}) [w^\pm z^0 H^0] 
     -  (\mg - \mH) c_{2W} 
       (C_{12} + C_{23})  [H^0 w^\pm w^\pm] \rt. \nn \\
&& \!\!\!\!\!\!\!\!\!\!\!\!  
     -  (\mg - \ma) (C_{12} + C_{23}) 
      [w^\pm H^0 A^0]  -  \mw 
      \lt( 2 C_0 + 2 C_{11} + C_{12} + C_{23} \rt)
    [W^\pm H^0 A^0] \nn \\
&& \!\!\!\!\!\!\!\!\!\!\!\!  
     -  \fr{c_{2W}}{c_W} 
       m_W^2 (- C_{12} + C_{23}) 
      [H^\pm H^0 Z^0]   +  \mw
     \lt( 2 C_0 - 2 C_{11} + 5 C_{12} + C_{23} \rt)
       [W^\pm Z^0 H^0]  \nn\\
&& \!\!\!\!\!\!\!\!\!\!\!\!  
    \lt.\lt.  +  c^2_W \mw 
     \lt( 4 C_{11} - 3 C_{12} - C_{23} \rt)
       [H^0 W^\pm W^\pm]
     +  \mw s^2_W   \lt( C_{23} - C_{12} \rt) 
        [H^0 w^\pm W^\pm]  - (H^0 \rightarrow h^0)\rt\}  \rt] \nn \\
&& \!\!\!\!\!\!\!\!\!\!\!\! 
+ \fr{4 N_c \mw}{16 \pi^2 v^2 c_W}
\lt[ \mb \tan \be \lt\{ ( - s_W^2 Q_b) (C_{12} - C_{11}) (tbb) 
   + (T_b - s_W^2 Q_b) (2 C_{23} + C_{12})(tbb)
\rt.\rt. \nn\\ 
&& \!\!\!\!\!\!\!\!\!\!\!\! 
\lt.      + (T_t - s_W^2 Q_t) (C_{12} + 2 C_{23})(tbb) \rt\}
      + \mt \cot \be \lt\{ (T_b - s_W^2 Q_b) 
  (C_0 + C_{11} + 2 C_{12} + 2 C_{23}) (tbb)\rt. \nn\\    
&& \!\!\!\!\!\!\!\!\!\!\!\! 
\lt.\lt.   
-  ( - s_W^2 Q_t) (C_{11} - C_{12}) (ttb) 
+  (T_t - s_W^2 Q_t) 
  (C_0 + C_{11} + 2 C_{12} + 2 C_{23})(ttb)\rt\} \rt]. 
\\
&& \!\!\!\!\!\!\!\!\!\!\!\! 
H^{HWZ (a)}(\mg,p_W^2,p_Z^2) =  \fr{4 N_c \mw}{16 \pi^2 v^2 c_W}
\lt[  \mb \tan \be ( - s_W^2 Q_b) 
\left\{ (C_{12} - C_{11}) (tbb)  
-  (T_b - s_W^2 Q_b)  C_{12}(tbb)      \rt. \rt. \nn\\ 
&& \!\!\!\!\!\!\!\!\!\!\!\! 
\lt. -  (T_t - s_W^2 Q_t)  C_{12}(ttb) \rt\}
+ \mt \cot \be \lt\{ -  (T_b - s_W^2 Q_b) (C_0 + C_{11}) (tbb)
 ( - s_W^2 Q_t) (C_{11} - C_{12}) (ttb)  \rt.\nn\\ 
&& \!\!\!\!\!\!\!\!\!\!\!\! 
\lt. \lt. -  (T_t - s_W^2 Q_t) (C_0 + C_{11})(ttb) \rt\}\rt].    
\end{eqnarray}
}
There is no contribution from the other diagrams to 
$G^{HWZ}$ and $H^{HWZ}$;
{\small 
\begin{eqnarray}
G^{HWZ (b)} = G^{HWZ (c)} 
= H^{HWZ (b,c)} = 0.
\end{eqnarray}
}

\subsubsection{The $H^+ W^- \gamma$ vertex}

\hspace*{18pt}
By making the similar decomposition to the $HWZ$ vertex, 
we obtain contributions of the $H^+ W^- \gamma$ vertex to each form factor.
{\small 
\begin{eqnarray}
&& \!\!\!\!\!\!\!\!\!\!\!\! 
  F^{HW\gamma (a)}(\mg,p_W^2,p_\gamma^2) = \fr{4 s_W}{16 \pi^2 v^2} 
\times \lt[ \tilde{K}_{\al\be} C_{24}[H^0H^\pm H^\pm] 
+ \tilde{L}_{\al\be} C_{24}[h^0H^\pm H^\pm]\rt. \nn \\
&& \!\!\!\!\!\!\!\!\!\!\!\!  
     + J_{\al\be}\lt\{   \fr{\mw}{2}  
     \lt( (p_\gamma^2 - p_W^2 ) C_0 
     - 2 p_\gamma \cdot (p_W C_{11} + p_\gamma C_{12})
       + p_W \cdot p_\gamma C_{23} + (D-1) C_{24} \rt)
      [H^0 W^\pm W^\pm] \rt. \nn \\
&& \!\!\!\!\!\!\!\!\!\!\!\! 
\lt.\lt.     +  \fr{\mw}{2} (\mg - \mH)  C_0[H^0 W^\pm w^\pm]
     -  \fr{\mw}{2} C_{24}[H^0 w^\pm W^\pm] - (\mg - \mH) 
 C_{24}[H^0 w^\pm w^\pm] \rt.\rt. \nn\\
&& \!\!\!\!\!\!\!\!\!\!\!\! 
\lt.\lt.- (H^0 \rightarrow h^0)\rt\}\rt]+ \fr{4 s_W N_c }{16 \pi^2 v^2} 
\lt[ \mb \tan \be \lt\{ Q_b 
\lt( \pw \cdot (\pw + \pg) C_{11} + \pg \cdot (\pw + \pg) C_{12} 
    + \pwsq C_{21} + \pgsq C_{22} \rt.\rt.\rt. \nn\\
&& \!\!\!\!\!\!\!\!\!\!\!\!
\lt.  + 2 \pw \cdot \pg C_{23}  + 4 C_{24} \rt)(tbb)
-  Q_b \lt( \pwsq  C_{11} + \pw \cdot \pg C_{12} 
+ \pwsq C_{21} + \pgsq C_{22} + 2 \pw \cdot \pg C_{23}
 + 2 C_{24} \rt)(tbb)\nn\\
&& \!\!\!\!\!\!\!\!\!\!\!\!
\lt.  - \mb \tan \be  Q_t
 \lt( \pgsq  C_{11} + \pg \cdot \pz C_{12} 
+ \pgsq C_{21} + \pwsq C_{22} + 2 \pw \cdot \pg C_{23} + 2 C_{24} 
        \rt)(ttb)+   \mt  Q_t  C_0(ttb) \rt\}   \nn\\
&& \!\!\!\!\!\!\!\!\!\!\!\! 
 + \mt \cot \be \lt\{ - Q_b
 \lt( (\pwsq + \pw \cdot \pg ) C_{0} 
     +  (2 \pwsq + \pw \cdot \pg ) C_{11} 
     + (p_\gamma^2 + 2 \pw \cdot \pg ) C_{12} 
+ \pwsq C_{21} + \pgsq C_{22}\rt.\rt. \nn\\
&& \!\!\!\!\!\!\!\!\!\!\!\! 
\lt.  + 2 \pw \cdot \pg C_{23} + 2 C_{24} 
        \rt)(tbb)   
+  \mb  Q_b  C_0(tbb) +  Q_t 
  \lt( \pg \cdot (\pw + \pg) C_{11} + \pw \cdot (\pw + \pg) C_{12} \rt. \nn\\
&& \!\!\!\!\!\!\!\!\!\!\!\! 
 \lt. + \pgsq C_{21} + \pwsq C_{22} + 2 \pw \cdot \pg C_{23} + 4 C_{24} 
        \rt)(ttb) -   Q_t
 \lt( (\pgsq + \pw \cdot \pg ) C_{0} 
     +  (2 \pgsq + \pw \cdot \pg ) C_{11}\rt. \nn\\ 
&&\!\!\!\!\!\!\!\!\!\!\!\! \lt.\lt. 
     + (\pwsq + 2 \pw \cdot \pg ) C_{12} 
+ \pgsq C_{21} + \pwsq C_{22} + 2 \pw \cdot \pg C_{23} + 2 C_{24} 
        \rt)(ttb) \rt].  \\ \nn \\
&& \!\!\!\!\!\!\!\!\!\!\!\! 
      F^{HW\gamma (b)}(\mg,p_W^2,p_\gamma^2) = 
\fr{4 s_W}{16 \pi^2 v^2} 
    \times \lt[  - \fr{1}{4} \tilde{K}_{\al\be} \rt.
        \lt\{  B_0 [H^0 H^\pm] + \fr{\mH - \mg}{\mg - \mw} 
        B_0 [H^0 H^\pm]   \rt.\nn \\
&& \!\!\!\!\!\!\!\!\!\!\!\! 
   \lt.  - \fr{p_\gamma^2 - p_W^2}{\mg - \mw} 
        (B_0 + 2 B_1) [H^0 H^\pm] \rt\} - \fr{1}{4} \tilde{L}_{\al\be} 
         \lt\{ B_0 [h^0 H^\pm]  + \fr{\mh - \mg}{\mg - \mw} 
        B_0 [h^0 H^\pm] \rt. 
  \nn \\
&& \!\!\!\!\!\!\!\!\!\!\!\! 
\lt.   - \fr{p_\gamma^2 - p_W^2}{\mg - \mw} 
        (B_0 + 2 B_1) [h^0 H^\pm]  \rt\}
    + J_{\al\be} \left\{ \fr{1}{4}  (\mg - \mH)  
          B_0 [H^0 w^\pm]   -    \fr{\mw}{2}   B_0 (p_W^2; W^\pm H^0)  
\right. \nn \\
&& \!\!\!\!\!\!\!\!\!\!\!\! 
     + \fr{1}{4}  \fr{m_W^2}{\mg - m_W^2} 
      \lt\{ \mg (B_0 - 2 B_1 + B_{21}) + D B_{22} \rt\}
      [H^0 W^\pm]   -  \fr{\mH}{4} \fr{\mH - \mg}{\mg - \mw} B_0 [H^0 w^\pm] 
  \nn\\
&& \!\!\!\!\!\!\!\!\!\!\!\! 
  \lt.\lt.     +  \fr{1}{2}  \fr{\mH - \mg}{\mg - \mw} (p_\gamma^2 - p_W^2) 
        (B_0 + 2 B_1) [H^0 w^\pm]
   +  \fr{\mw}{2}  \fr{p_\gamma^2 - p_W^2}{\mg - \mw} 
        (B_0 - B_1) [H^0 W^\pm]  - (H^0 \rightarrow h^0) \rt\}\rt]
\nn \\
&& \!\!\!\!\!\!\!\!\!\!\!\!  
  + \fr{4 s_W}{16 \pi^2 v^2} \left[ 
  - \frac{\pgsq - \pwsq}{\mg - \mw}   \left\{ \mb \tan \be B_1    
+ \mt \cot \be (B_1 + B_0) \right\}[tb] 
\rt. \nn \\
&& \!\!\!\!\!\!\!\!\!\!\!\! 
\lt.- \frac{1}{\mg - \mw} 
\lt\{ (\mb \tan \be - \mt \cot \be) (\mg (B_1 + B_{21}) 
  + D B_{22})
         + \mt \mb (\tan \be - \cot \be) B_0 \rt\}[tb] \rt]. \\
\nn\\   
&& \!\!\!\!\!\!\!\!\!\!\!\! 
     F^{HW\gamma (c)}(\mg,p_W^2,p_\gamma^2) = - 
\fr{s_W}{16 \pi^2 v^2} \frac{1}{\mg - \mw} 
 \lt\{  \tilde{\Pi}^B_{H w} - T_1 + ( \pgsq - p_W^2 + m_W^2 ) T_2 \rt\}.
\end{eqnarray}
}
where $T_1$ and $T_2$ and
$\tilde{\Pi}_{Hw}^B$ 
are defined in \eq{tadB}, \eq{tadA} and 
\eq{two}.
{\small 
\begin{eqnarray}
&& \!\!\!\!\!\!\!\!\!\!\!\! 
  G^{HW\gamma (a)}(\mg,p_W^2,p_\gamma^2) = 
\fr{4 \mw s_W}{16 \pi^2 v^2} 
    \lt[ \tilde{K}_{\al\be}  
            (C_{12} + C_{23})[H^0H^\pm H^\pm] 
     +  \tilde{L}_{\al\be} 
            (C_{12} + C_{23})[h^0H^\pm H^\pm]\rt. \nn \\
&& \!\!\!\!\!\!\!\!\!\!\!\!  
     + J_{\al\be} \lt\{  \fr{\mw}{2}  
     \lt( 4 C_{11} - 3 C_{12} - C_{23} \rt)
      [H^0 W^\pm W^\pm]     
     +  \fr{\mw}{2} 
     ( C_{12} - C_{23} )[H^0 w^\pm W^\pm] 
       \rt. \nn\\
&& \!\!\!\!\!\!\!\!\!\!\!\! 
\lt.\lt.     -  (\mg - \mH) (C_{12} + C_{23}) [H^0 w^\pm w^\pm] 
    - (H^0 \rightarrow h^0)   \rt\} \rt] 
+ \fr{4 \mw s_W N_c}{16 \pi^2 v^2} 
\lt[ \mb \tan \be Q_b ( C_{12} - C_{11})(tbb) \rt. \nn\\
&& \!\!\!\!\!\!\!\!\!\!\!\! 
+ \mb \tan \be Q_b (2 C_{23} +  C_{12})(tbb) + \mt \cot \be Q_b
 ( C_{0} + C_{11} + 2 C_{12} + 2 C_{13} )(tbb) + \mt \cot \be Q_t 
  \nn\\
&& \!\!\!\!\!\!\!\!\!\!\!\! \lt.
  \times ( C_{12} - C_{11} ) (ttb)+ \mb \tan \be  Q_t
 ( 2 C_{23} +  C_{12} ) (ttb)  
+ \mt \cot \be Q_t ( C_{0} + C_{11} + 2 C_{12} + 2 C_{23} )(ttb)  \rt].\\
&&\nn\\
&& \!\!\!\!\!\!\!\!\!\!\!\! 
  H^{HW\gamma (a)}(\mg,p_W^2,p_\gamma^2) = 
 \fr{4 \mw s_W N_c}{16 \pi^2 v^2} 
\lt[ \mb \tan \be Q_b ( C_{12} - C_{11})(tbb) 
- \mb \tan \be Q_b C_{12}(tbb) \rt. \nn\\
&& \!\!\!\!\!\!\!\!\!\!\!\!  - \mt \cot \be Q_b
 ( C_{0} + C_{11} )(tbb)   + \mt \cot \be Q_t 
  ( C_{11} - C_{12} ) (ttb) - \mb \tan \be  Q_t
 C_{12} (ttb)   \nn\\
&& \!\!\!\!\!\!\!\!\!\!\!\! \lt.
- \mt \cot \be Q_t ( C_{0} + C_{11} )(ttb)  \rt].
\end{eqnarray}
}
and 
{\small 
\begin{eqnarray}
G^{HW\gamma (b,c)} = H^{HW\gamma (b,c)} = 0.
\end{eqnarray}
}

\subsection{Tadpole diagrams and the $w$-$H$ two point function}

\hspace*{18pt}
The tadpole graphs $i\, T_H$ and $i\, T_h$ are calculated as  
{\small 
\begin{eqnarray}
&& \!\!\!\!\!\!\!\!\!\!\!\! 
 T_H = \fr{1}{16 \pi^2 v}
\lt[  \mH \cs \ab \lt( A[w^\pm] + \fr{1}{2}A[z^0] \rt) \rt.\nn\\
&& \!\!\!\!\!\!\!\!\!\!\!\! 
 + \lt\{ \lt( \fr{\cs \al \sn^2 \be}{\cs \be} - 
                                   \fr{\sn \al \cs^2 \be}{\sn \be} \rt) \mH 
  + 2 \cs \ab  \mg 
             + \fr{\sin (\al + \be)}
                  {\sin \be \cos \be} M^2\rt\} 
  A[H^\pm]  \nn \\
&& \!\!\!\!\!\!\!\!\!\!\!\! 
 + \lt\{  \lt( \fr{\cs \al \sn^2 \be}{\cs \be} - 
                                   \fr{\sn \al \cs^2 \be}{\sn \be} \rt) \mH
  + 2 \cs \ab \ma  
             + \fr{\sin (\al + \be)}
                  {\sin \be \cos \be} M^2\rt\} \fr{1}{2}A[A^0]   \nn \\
&& \!\!\!\!\!\!\!\!\!\!\!\! 
      + \fr{3}{2} \lt\{ 
                         \lt( \fr{\cs^3 \al}{\cs \be} + 
                              \fr{\sn^3 \al}{\sn \be} \rt) \mH 
        - \fr{\cos 2 \be}{\cos \be \sin \be}\sin (\al - \be) M^2 
                    \rt\}  A[H^0] \nn\\ 
&& \!\!\!\!\!\!\!\!\!\!\!\! 
      + \lt\{ \fr{1}{2} ( \mH + 2 \mh ) \fr{\sn 2\al}{\sn 2\be}  
                - \fr{M^2}{4 \cos \be \sin \be}  
                (-3 \sin 2 \al + \sin 2 \be) \rt\}
                                           \cos (\al - \be) A[h^0] \nn \\
&& \!\!\!\!\!\!\!\!\!\!\!\! 
   \lt.   + 8 \cs \ab  \lt( \mw A[W^\pm] + 
                          \fr{1}{2} \mz A[Z^0] \rt)  
    - 4 N_c \left( \frac{\cos \al}{\cos \be}  A[b]   
        + \frac{\sin \al}{\sin \be}  A[t] \rt)\rt],  \label{tad1}\\
&& \!\!\!\!\!\!\!\!\!\!\!\! 
    T_h =  \fr{1}{16 \pi^2 v} 
\lt[ - \mh \sn \ab \lt( A[w^\pm] + \fr{1}{2}A[z^0] \rt) \rt.\nn\\
&& \!\!\!\!\!\!\!\!\!\!\!\! 
 + \lt\{ \lt( \fr{\sn \al \sn^2 \be}{\cs \be} - 
                                   \fr{\cs \al \cs^2 \be}{\sn \be} \rt) \mh
                 - 2  \sn \ab  \mg 
             + \fr{\cos (\al + \be)}
                  {\sin \be \cos \be} M^2 \rt\}  A[H^\pm] \nn \\
&& \!\!\!\!\!\!\!\!\!\!\!\! 
 + \lt\{\lt( \fr{\sn \al \sn^2 \be}{\cs \be} - 
                                   \fr{\cs \al \cs^2 \be}{\sn \be} \rt) \mh 
                 - 2 \sn \ab  \ma  
             + \fr{\cos (\al + \be)}
                  {\sin \be \cos \be} M^2 \rt\} 
\fr{1}{2}A[A^0]   \nn \\
&& \!\!\!\!\!\!\!\!\!\!\!\! 
 - \fr{3}{2} \lt\{ \lt( \fr{\sn^3 \al}{\cs \be} - 
                              \fr{\cs^3 \al}{\sn \be} \rt) \mh
        +\fr{\cos 2 \be}{\cos \be \sin \be}\cos (\al - \be) M^2 
                    \rt\} A[h^0]  \nn\\
&& \!\!\!\!\!\!\!\!\!\!\!\! 
 + \fr{1}{2} \lt\{ ( 2 \mH + \mh ) \fr{\sn 2\al}{\sn 2\be} 
               - \fr{M^2}{4 \cos \be \sin \be}  
                (3 \sin 2 \al + \sin 2 \be) \rt\}\sn \ab A[H^0] \nn \\
&& \!\!\!\!\!\!\!\!\!\!\!\! 
 \lt. - 8 \sn \ab \lt( \mw A[W^\pm] + \fr{1}{2} \mz A [Z^0] \rt)
    - 4 N_c \left( \frac{\sin \al}{\cos \be}  A[b]   
                + \frac{\cos \al}{\sin \be}  A[t] \rt) \rt].  
\label{tad2}
\end{eqnarray}

{\normalsize
The $w$-$H$ two point function is given by} 
\begin{eqnarray}
   \Pi_{wH}(p^2) = \Pi_{wH}^A(p^2) + \Pi_{wH}^B + \Pi_{wH}^C, \label{pigw}
\end{eqnarray}
}
where $\Pi_{wH}^B$ is the contribution of the diagrams which 
include a quartic Higgs-self coupling constants 
and $\Pi_{wH}^C$ is the tadpole contributions. The explicit formulas are 
{\small 
\begin{eqnarray}
&& \!\!\!\!\!\!\!\!\!\!\!\!  \Pi_{H w}^A(\mg) = \fr{1}{16 \pi^2 v^2}
\lt[  (m_{H^0}^2-m_{H^\pm}^2) \tilde{K}_{\al\be} 
      B_0 [H^0 H^\pm] + (m_{h^0}^2-m_{H^\pm}^2)  \tilde{L}_{\al\be} 
          B_0 [h^0 H^\pm]   + J_{\al\be} \left\{ -m_W^2 \rt.        \rt.\nn \\
&& \!\!\!\!\!\!\!\!\!\!\!\! 
 \lt. \times \lt( p^2 (B_0 - 2 B_1 + B_{21}) + D B_{22} \rt) 
     [H^0 W^\pm]   +   \mH (\mH - \mg)  B_0 [H^0 w^\pm] 
     - (H^0 \rightarrow h^0)\rt\} \nn \\
&& \!\!\!\!\!\!\!\!\!\!\!\! 
 + \frac{4 N_c}{16 \pi^2 v^2} 
\lt[  (\mb \tan \be - \mt \cot \be) 
        \lt( \mg (B_{1} + B_{21}) + D B_{22} \rt) [tb]
    - \mt \mb (\tan \be - \cot \be) B_0 [tb]   \rt], \nn\\\\ 
&& \!\!\!\!\!\!\!\!\!\!\!\!  \Pi_{H w}^B= \fr{1}{16 \pi^2 v^2} {
    \tilde{\Pi}_{H w}^B}=
\fr{1}{16 \pi^2 v^2} 
      \lt[ 2(\mH - \mh) J_{\al\be} 
      \lt( A[W^\pm] + \fr{1}{4} A[Z^0]\rt) \rt. \nn \\
&& \!\!\!\!\!\!\!\!\!\!\!\! 
2 \lt\{  +  (K_{\al\be} - J_{\al\be})\mH 
+  (L_{\al\be} + J_{\al\be}) \mh      
      - 2 \cot 2 \be \;M^2\; \rt\} \lt( A[H^\pm] + \frac{1}{4} A[A^0]\rt) 
 \nn\\
&&\!\!\!\!\!\!\!\!\!\!\!\!
+ J_{\al\be} \mg \lt( A[h^0] - A[H^0] \rt)+ \fr{1}{4} \sn 2 \be 
     \lt( \fr{\sn^4 \al}{\sn^2 \be} - \fr{\cs^4 \al}{\cs^2 \be} 
          + \fr{\sn 2 \al \cs 2 \al}{\sn 2 \be} \rt) \mH A[H^0] \nn\\
&&\!\!\!\!\!\!\!\!\!\!\!\! 
+ \fr{1}{4} \sn 2 \be 
\lt(  \fr{\cs^4 \al}{\sn^2 \be} - \fr{\sn^4 \al}{\cs^2 \be} 
     + \fr{\sn 2\al \cs 2\al}{\sn 2 \be}  \rt) \mh A[h^0]\nn\\
&& \!\!\!\!\!\!\!\!\!\!\!\!+ \fr{1}{4} \sn 2 \be 
     \lt(  \fr{\sn^2 \al \cs^2 \al}{\sn^2 \be} 
         - \fr{\sn^2 \al \cs^2 \al}{\cs^2 \be} 
         - \fr{\sn 2\al \cs 2 \al}{\sn 2\be} \rt)  
   \lt( \mh A[H^0] + \mH A[h^0] \rt) \nn \\
&&\!\!\!\!\!\!\!\!\!\!\!\! \lt. 
- \fr{M^2}{2}\fr{\cos 2 \be}{\cos \be \sin \be} 
    \lt(\sin^2 (\al - \be) A[H^0] +  \cos^2 (\al - \be) A[h^0] \rt)
 \rt] , \label{two}\\
&& \!\!\!\!\!\!\!\!\!\!\!\!  \Pi_{H w}^C= 
 \frac{1}{v} \left(  - T_1 + \mg T_2 \right), 
\end{eqnarray}
where
\begin{eqnarray}
  T_1 &=& 16 \pi^2 v^2 \left\{
 \sn \ab T_H + \cs \ab T_h \right\}, \label{tadB}\\
  T_2 &=& 16 \pi^2 v^2 \left\{
\fr{1}{\mH} \sn \ab T_H + \fr{1}{\mh} \cs \ab T_h \right\}.\label{tadA}
\end{eqnarray}
}

\subsection{The $t$ channel contribution}

\hspace*{18pt}
The contribution of the t-channel diagram (Fig.~1(b)) 
is only from the $W^+H^-$ mixing.
When we write the $W^\mu H$ two-point function as 
\begin{eqnarray}
     i \Pi_{{W H}}^\mu(p) = i p^\mu \Pi_{WH}(p^2),
\end{eqnarray}
the contribution to the form factor is expressed as 
\begin{eqnarray}
  F_{i,\tau}^{t}(t) = \delta_{i, 1} \delta_{\tau, -1} 
                \frac{g^2}{2} \frac{1}{m_{H^\pm}^2 - m_W^2} 
                \Pi_{WH}(m_{H^{\pm}}^2). 
\end{eqnarray}
{\small where
\begin{eqnarray}
  \Pi_{WH}(p^2) &=& 
  \frac{m_W}{16 \pi^2 v^2}  \left[ 
\tilde{K}_{\al\be} \left( 2 B_1 + B_0 \right)[H^0 H^{\pm}]  
 + \tilde{L}_{\al\be} \left( 2 B_1 + B_0 \right)[h^0 H^{\pm}]
\rt. \nn \\
 && \!\!\!\!\!\!\!\!\!\!\!\! \!\!\!\!\!\!\!\!\!\!\!\! 
+ J_{\al\be} \lt\{ 2 m_W^2  (B_0 - B_1)[H^0 W^\pm] 
 + (\mH - \mg)   \left( 2 B_1 + B_0 \right)[H^0 w^\pm] 
 - (H^0 \rightarrow h^0)\rt\} \nn\\
 && \!\!\!\!\!\!\!\!\!\!\!\! \!\!\!\!\!\!\!\!\!\!\!\! 
 \lt. - 4 N_c \lt\{ \mb \tan \be B_1 + \mt \cot \be (B_1 + B_0) \rt\}
        [tb]  - T_2 \rt] , \label{pi-WH}
\end{eqnarray}
where the tadpole contribution $T_2$ is given in \eq{tadA}.
}

\subsection{The box-diagram}

\hspace*{18pt}
The contribution from the box diagrams (Fig.~1(c)) is parametrized as  
\begin{eqnarray}
  F_{i,\tau}^{\rm box}(s,t) = - \fr{1}{16 \pi^2}
         \frac{g^4}{4} m_W J_{\al\be} 
        \left\{ f_i^{\rm box}[\nu,W,H^0,W] - 
                f_i^{\rm box}[\nu,W,h^0,W]   \right\} \delta_{\tau, -1}.
\end{eqnarray}
The functions $f_i^{\rm box}$ are calculated as 
{\small
\begin{eqnarray}
  f_1^{\rm box} [\nu, W, S, W] 
                 &=& \lt\{ 2 (t - m_{H^\pm}^2) D_{11} 
                  + 2 m_{H^\pm}^2 D_{12} 
                  + (s - m_{H^\pm}^2 - m_W^2) D_{13} \rt. \nn \\
&&                \lt. 
                  + m_{H^\pm}^2 D_{22} 
                  + m_W^2 D_{23}  +  (t - m_{H^\pm}^2) D_{24} 
        + (- s - t + m_{H^\pm}^2) D_{25} \rt. \nn \\
&&                \lt.           
                  + (s - m_{H^\pm}^2 - m_W^2) D_{26} 
                  + 4 D_{27} \rt\} [\nu, W, S, W], \\ 
  f_2^{\rm box}[\nu, W, S, W] &=& m_W^2 D_{13}[\nu, W, S, W] , \\
  f_3^{\rm box}[\nu, W, S, W] 
        &=& m_W^2 (\frac{1}{2} D_{11} + D_{13})[\nu, W, S, W] , 
\end{eqnarray}
where 
\begin{eqnarray}
  D_{ij}[\nu, W, S, W] 
= D_{ij} (k^2, p_H^2, p_W^2, \overline{k}^2; 0, m_W, m_S, m_W), \;\;
(S = h^0,  H^0). 
\end{eqnarray}
}

\subsection{Finite renormalization effects}

\hspace*{18pt}
The counterterm in Eq.~\eq{fdec} is obtained in terms of 
${\rm Re} \left(\Pi_{HW}(m_{H^\pm}^2)\right)$ and 
${\rm Re} \left(\Pi_{Hw}(m_{H^\pm}^2)\right)$.  
We decompose $\delta F_{i,\tau}$ into three parts as similarly to 
the one-loop diagram part in Eq.~\eq{fdec},
{\small 
\begin{eqnarray}
  \delta F_{i,\tau}(s,t)=\delta F_{i,\tau}^Z(s)+
                    \delta F_{i,\tau}^\gamma(s)+\delta F_{i,\tau}^t(t).  
\end{eqnarray}
where each part in RHS is written 
\begin{eqnarray}
  \delta F_{i,\tau}^V(s) &=& \delta_{i,1}
              g m_W^{} C_V \frac{1}{s-m_V^2} \delta F^{HWV}
                (m_W^2,s,m_{H^\pm}^2),\\
  \delta F_{i,\tau}^t(t) &=& -  
                \delta_{i,1} \delta_{\tau, -1} 
                         \frac{g^2}{2(m_{H^\pm}^2-m_W^2)} 
            {\rm Re} \left( \Pi_{WH} (m_{H^\pm}^2) \right) .
\end{eqnarray}
where $V$ represents $Z$ or $\gamma$, and $F^{HWV}(m_W^2,m_Z^2)$ and 
$F_{i,\tau}^t$ are expressed by 
\begin{eqnarray}
\delta F^{HWZ}(p_W^2,p_Z^2,m_{H^\pm}^2) \!\!\!\!&=&\!\!\!\! \frac{1}{c_W}
    \left(  c_W^2 \frac{p_W^2-p_Z^2}{m_{H^\pm}^2-m_W^2}
           - s_W^2 \right) \frac{1}{m_W^{}} 
                  {\rm Re} \left( \Pi_{WH} (m_{H^\pm}^2) \right)\nonumber\\
&&- \frac{1}{c_W}\frac{s_W^2 m_{H^\pm}^2}{m_{H^\pm}^2-m_W^2} 
                  {\rm Re} \left( \Pi_{wH} (m_{H^\pm}^2) \right), \\    
\delta F^{HW\gamma}(p_W^2,p_\gamma^2,m_{H^\pm}^2) \!\!\!\!&=& \!\!\!\!s_W 
             \left( 1 + \frac{p_W^2-p_\gamma^2}{m_{H^\pm}^2-m_W^2}
             \right) \frac{1}{m_W^{}} 
                  {\rm Re} \left( \Pi_{WH} (m_{H^\pm}^2) \right)\nonumber\\
&&    +  \frac{s_W}{m_{H^\pm}^2-m_W^2}
             {\rm Re} \left( \Pi_{wH} (m_{H^\pm}^2) \right), 
\end{eqnarray}
where $\Pi_{wH} (p^2)$ and $\Pi_{WH} (p^2)$ are 
given in Eqs.~\eq{pigw} and \eq{pi-WH}.
}

\newpage

\newpage

\section*{TABLE CAPTION}
\begin{description}
\item[Table 1:] The list of the kinematical factors  
$K_{i,\tau}(k,\overline{k},\overline{\lambda})$. 
\end{description}              

\section*{FIGURE CAPTIONS}
\begin{description}
\item[Fig. 1:] The diagrams for $e^+e^- \rightarrow H^-W^+$. The circles 
              in $(a)$, and $(b)$ represent all one-loop diagrams relevant 
              to the $HWV$ vertices ($V=\gamma, Z^0$) and the $HW$ mixing.  
              The arrows on the $H^\pm$ bosons and the $W$ boson lines 
              indicate the flow of negative electric charge.  

\item[Fig. 2:] The $HWV$ vertices ($V=\gamma, Z^0$). 
              The arrows on the $H^\pm$ boson and the $W$ boson lines indicate
              the flow of negative electric charge.  
              
\item[Fig. 3:] The total cross section of $e^+e^- \rightarrow H^-W^+$ 
               for $m_{H^\pm} = 200$ GeV at $\sqrt{s} = 500$ GeV as a 
               function of $\tan \beta$ in the 2HDM (solid lines) 
               and in the MSSM (dashed line). 
               For the 2HDM, three solid curves 
               correspond to $m_{A^0} = 300$, $600$ and $1200$ GeV. 
               The other parameters are chosen as $\al = \be - \pi/2$, 
               $m_{h^0} = 120$ GeV, $m_{H^0} = 210$ GeV and $M=0$ GeV.     

\item[Fig. 4:] The $\sqrt{s}$ dependence of the total cross section 
               of $e^+e^- \rightarrow H^-W^+$ for $m_{H^\pm} = 200$ GeV 
               for various $\tan\be$ in the non-SUSY 2HDM. 
               Solid curves are $\tan\be=0.3, 0.5, 1, 2, 4$ and 
               dotted curves are $\tan\be=8, 16, 32$.
               The other parameters are chosen as $\al = \be - \pi/2$, 
               $m_{h^0} = 80$ GeV, $m_{H^0} = 210$ GeV, $m_{A^0} = 1200$ GeV 
               and $M=0$ GeV.  

\item[Fig. 5:] The upper bound of the cross section of 
               $e^+e^- \rightarrow H^-W^+$ for $m_{H^\pm} = 200$ GeV 
               at $\sqrt{s} = 500$ GeV as a function of $\tan\be$ 
               under the conditions \eq{uniyuk} and \eq{unilam} 
               in the non-SUSY 2HDM (solid curve).  
               The dotted curve represent the cross section where  
               the condition \eq{unilam} is switched off.
               The dashed curve represent the cross section where 
               only $t$-$b$ loop contributions are included.                
                             
\item[Fig. 6:] The possible enhancement of the total cross section 
               of $e^+e^- \rightarrow H^-W^+$ for various $m_{H^\pm}$ 
               at $\sqrt{s} = 500$ GeV as a function of $\tan\be$ 
               in the non-SUSY 2HDM under the conditions \eq{uniyuk} 
               and \eq{unilam}. 

\item[Fig. 7(a)]
  The first group of the Feynman diagrams (the 't Hooft-Feynman gauge) 
  of the $HWV$ vertices 
  ($V=\gamma,Z^0$), which correspons to $X^{HWV(a)}$ ($X=F,G$ an $H$) 
  in Appendix A.~1.

\item[Fig. 7(b)]
  The second group of the Feynman diagrams (the 't Hooft-Feynman gauge) 
  of the $HWV$ vertices 
  ($V=\gamma,Z^0$), which correspons to $X^{HWV(b)}$ ($X=F,G$ an $H$) 
  in Appendix A.~1.

\item[Fig. 7(c)]
  The third group of the Feynman diagrams (the 't Hooft-Feynman gauge) 
  of the $HWV$ vertices 
  ($V=\gamma,Z^0$), which correspons to $X^{HWV(c)}$ ($X=F,G$ an $H$) 
  in Appendix A.~1.
               
\end{description}

\newpage 

\begin{center}
\vspace*{6mm}
\begin{tabular}{|c|c|c|c|}\hline
   &  $K_{1,\tau}(k,\overline{k},\overline{\lambda})$ & 
      $K_{2,\tau}(k,\overline{k},\overline{\lambda})$ & 
      $K_{3,\tau}(k,\overline{k},\overline{\lambda})$ \\ \hline
     $\overline{\lambda} = 0$ & 
     $- \frac{1}{2 m_W} (s - m_{H^{\pm}}^2 + m_W^2)  \sin \Theta$ & 
     $\frac{1}{2} \frac{s^2}{m_W^3} \beta_{HW}^2 \sin \Theta$ &
     $0$                      \\ \hline
     $\overline{\lambda} = \pm$ & 
     $\sqrt{\frac{s}{2}} (\mp \cos \Theta + \tau)$ & 
     $0$ &
     $- \frac{s}{m_W^2} \sqrt{\frac{s}{2}} \beta_{HW} 
     (\cos \Theta \mp \tau)$                      \\ \hline
\end{tabular} 

\vspace{1.2cm}
{\bf Table 1}
\end{center}

\newpage
\begin{minipage}[t]{15cm}  
\epsfxsize=15cm
\epsfbox{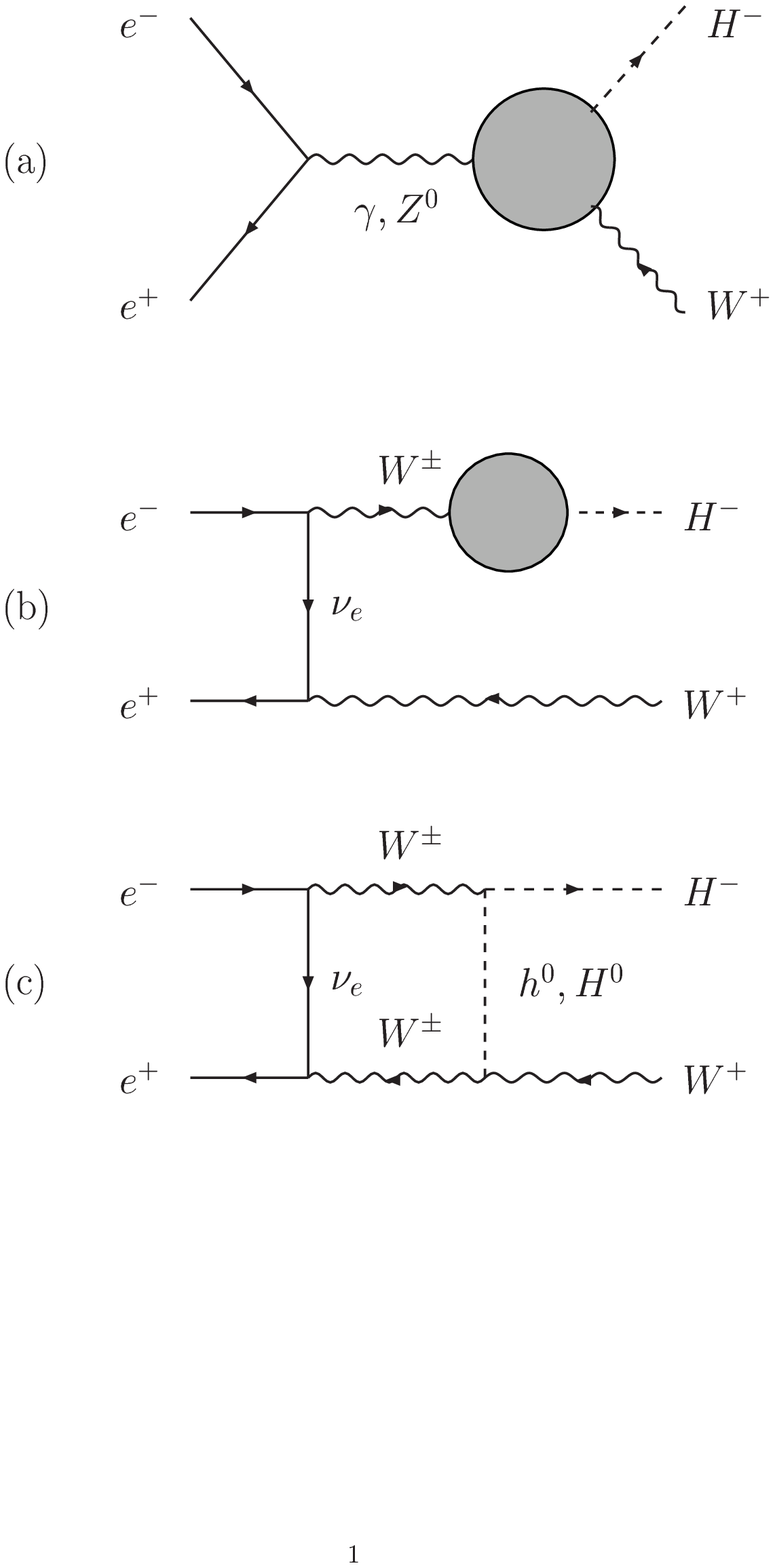} 
\begin{center}
  {\bf Figure 1} 
\end{center}
\end{minipage}

\begin{minipage}[t]{15cm}  
\epsfxsize=16cm
\epsfbox{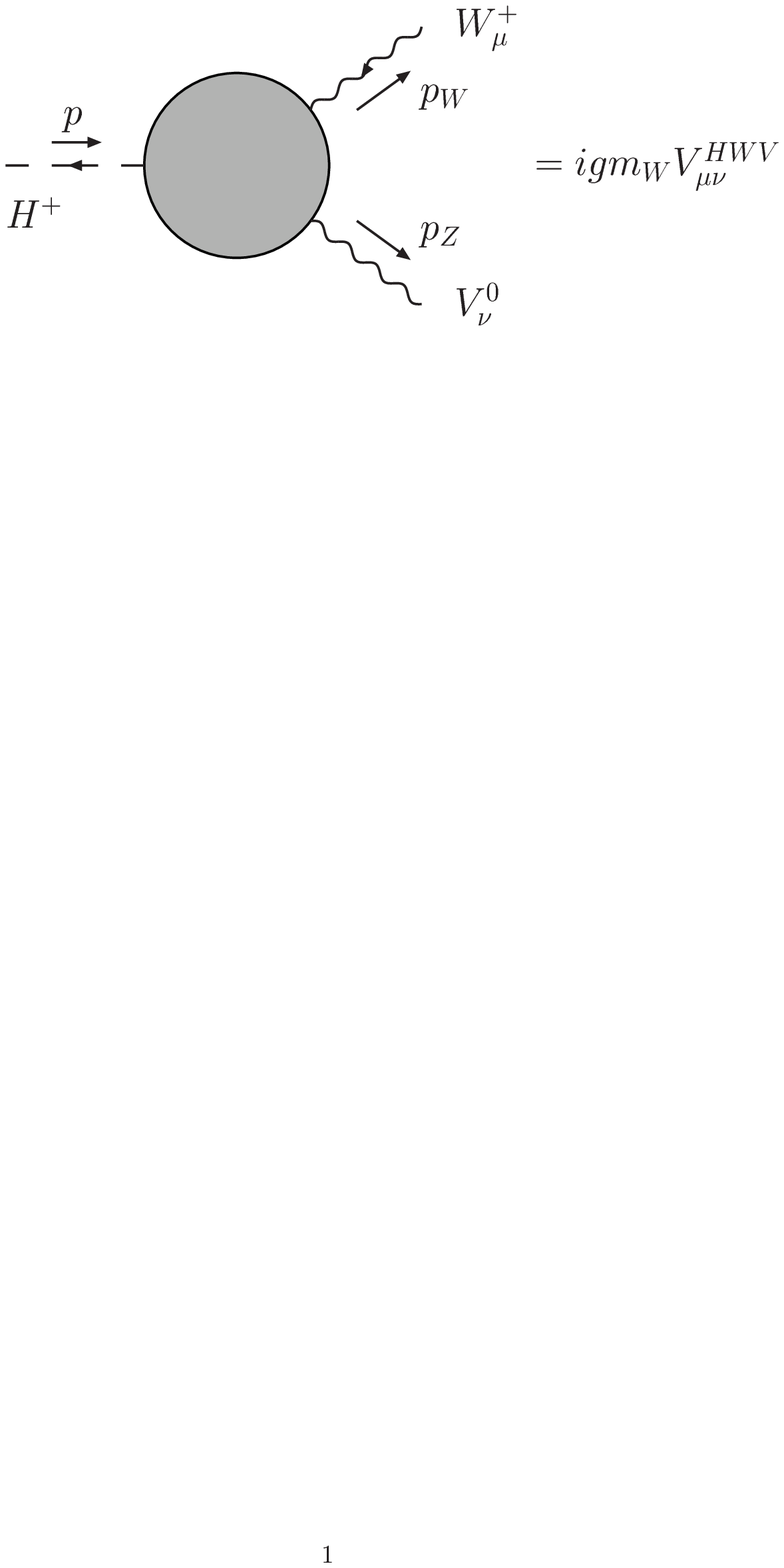}
\begin{center}
  {\bf Figure 2} 
\end{center}
\end{minipage}

\begin{minipage}[t]{15cm}  
\epsfxsize=15cm
\epsfbox{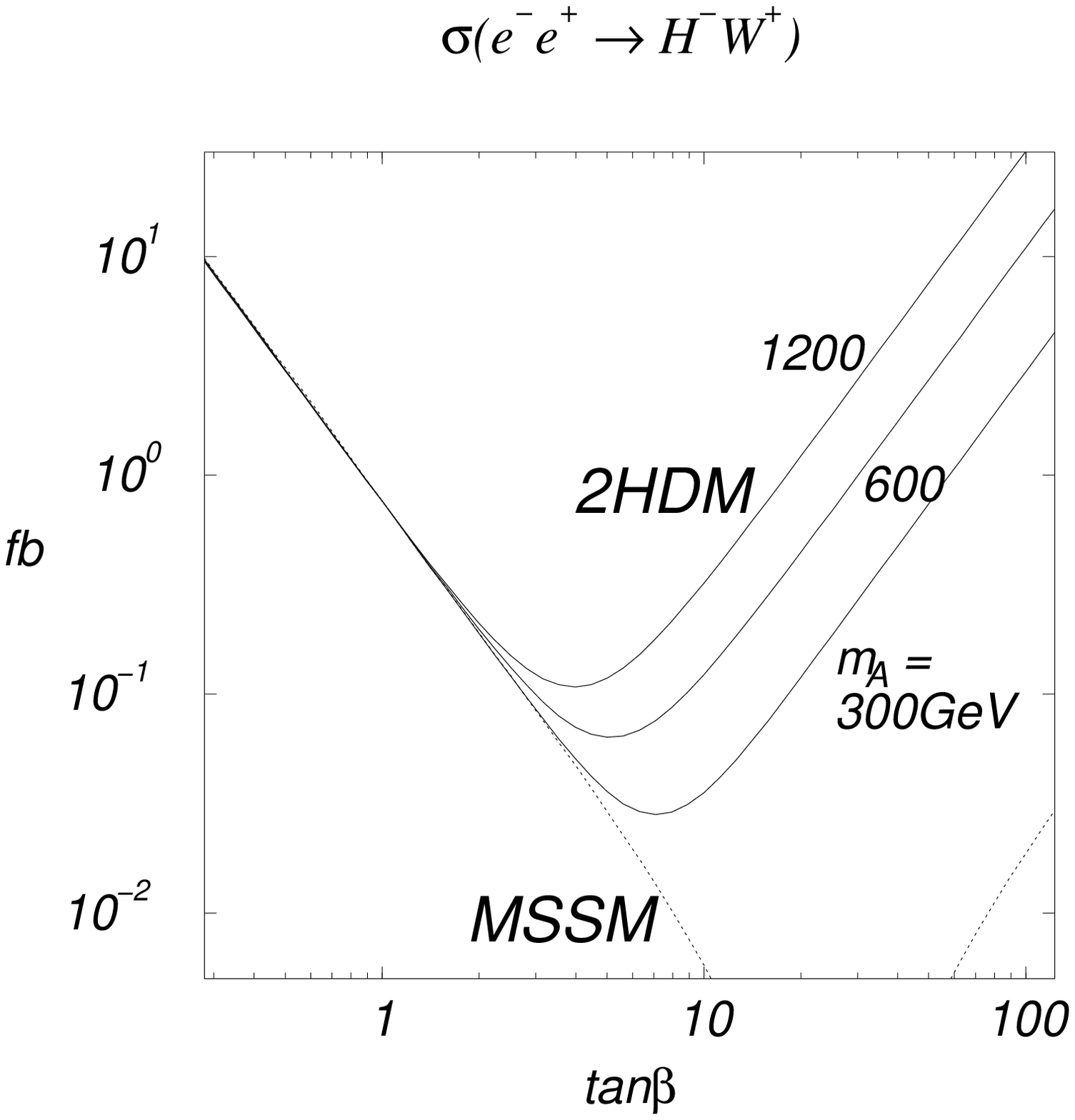}
\begin{center}
  {\bf Figure 3} 
\end{center}
\end{minipage}

\begin{minipage}[t]{15cm}  
\epsfxsize=15cm
\epsfbox{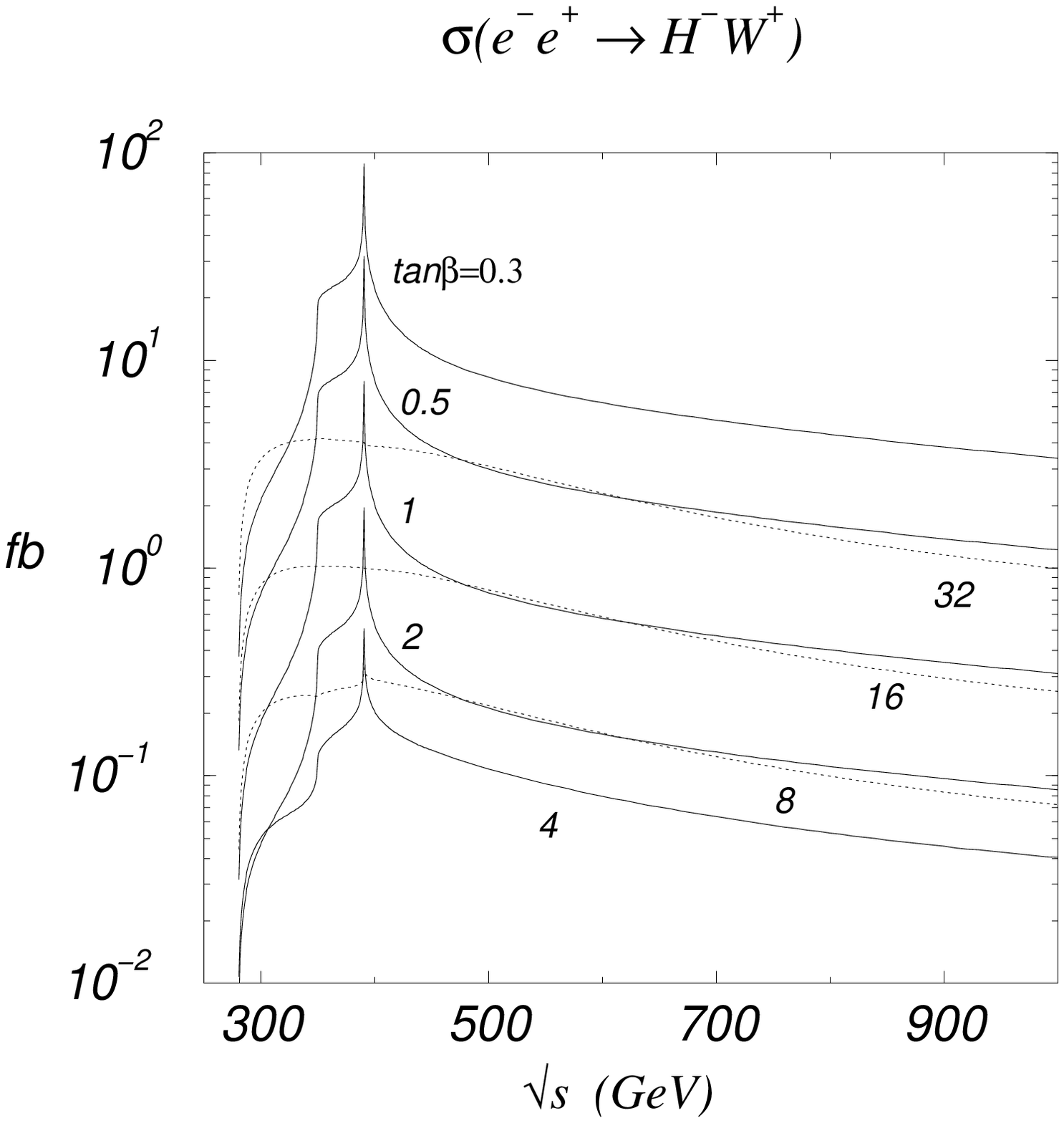}
\begin{center}
  {\bf Figure 4} 
\end{center}
\end{minipage}

\begin{minipage}[t]{15cm}  
\epsfxsize=15cm
\epsfbox{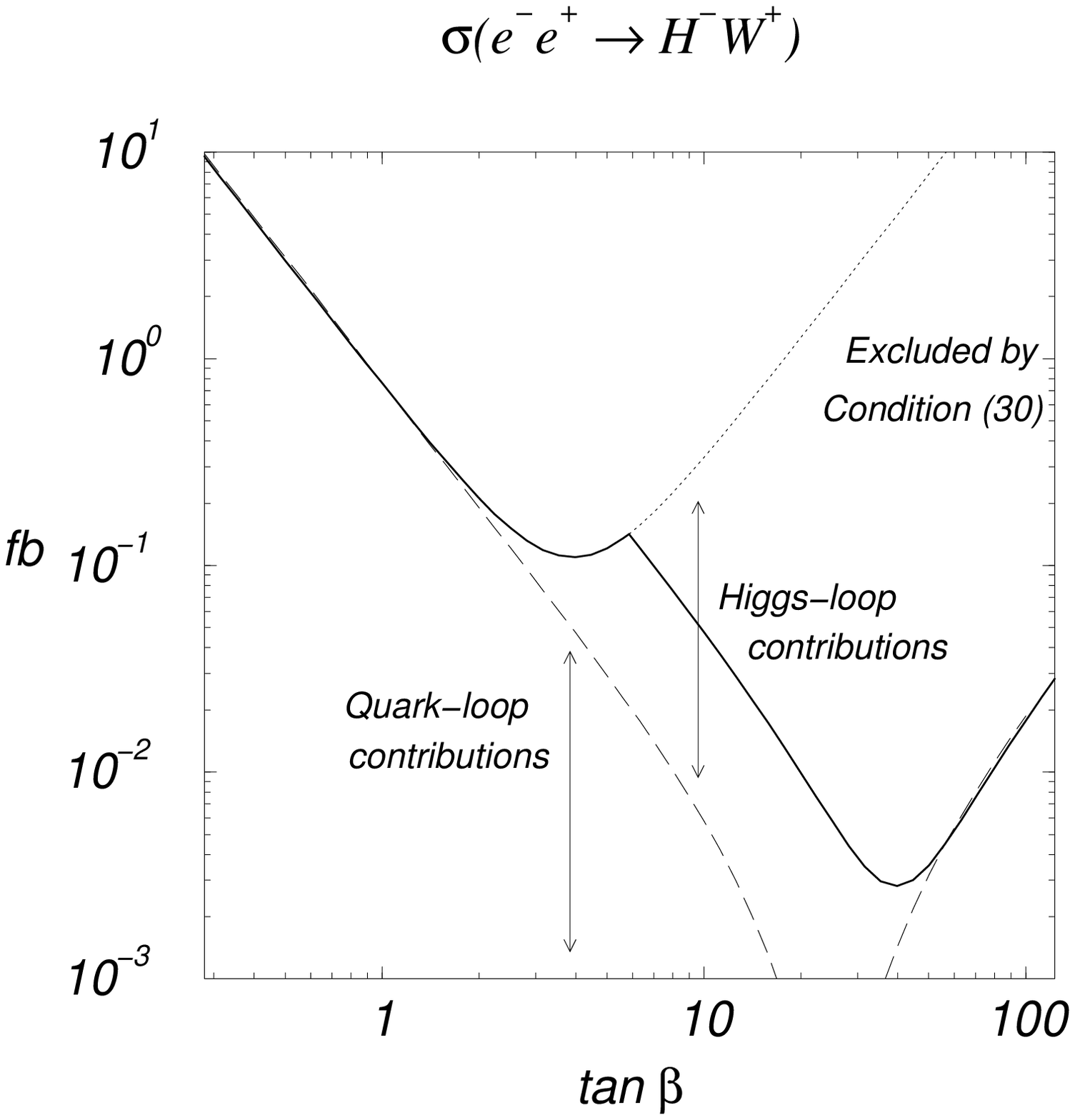}
\begin{center}
  {\bf Figure 5} 
\end{center}
\end{minipage}

\begin{minipage}[t]{15cm}  
\epsfxsize=15cm
\epsfbox{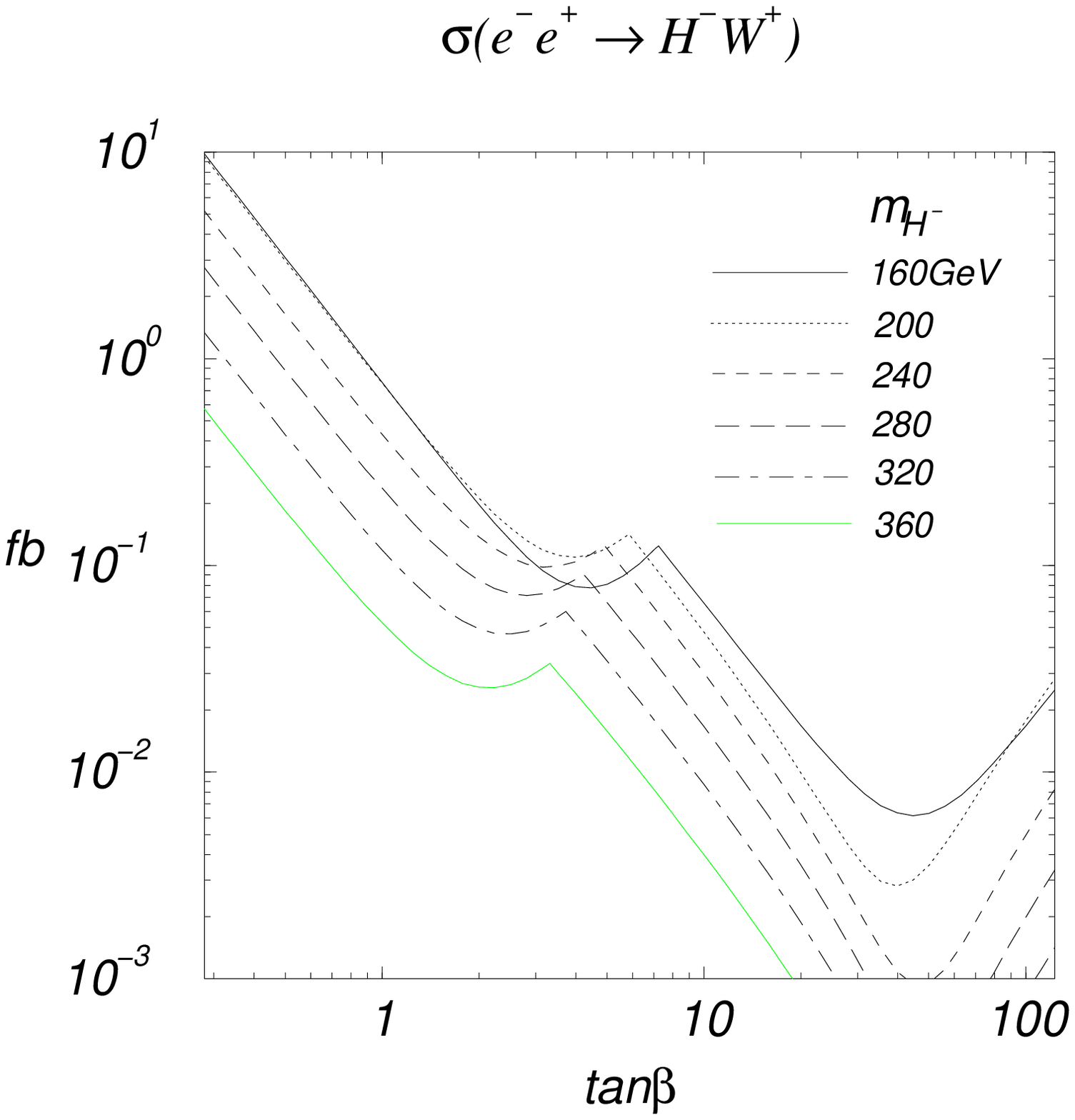}
\begin{center}
  {\bf Figure 6} 
\end{center}
\end{minipage}

\begin{minipage}[t]{15cm}  
\epsfxsize=15cm
\epsfbox{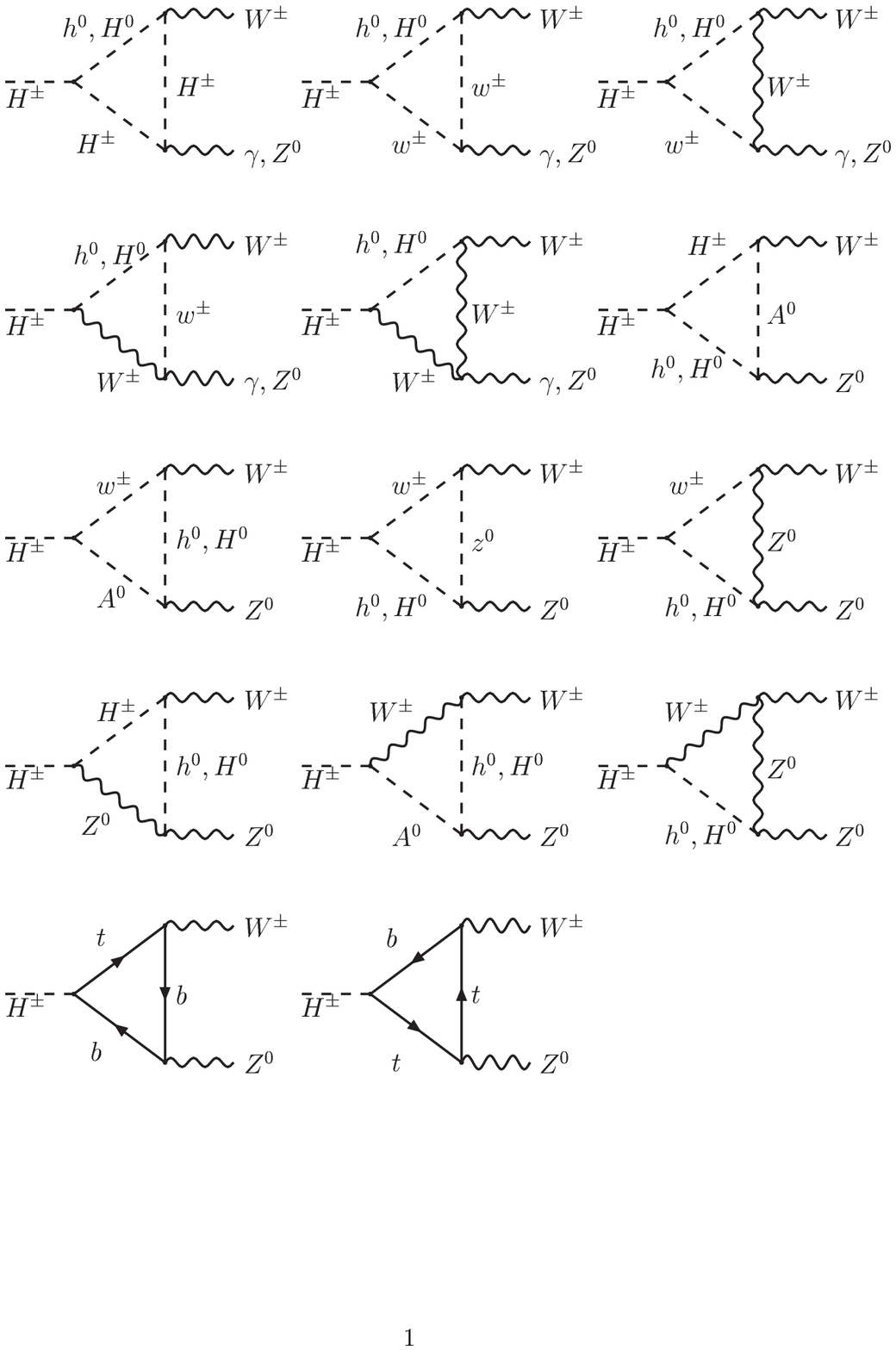}
\begin{center}
  {\bf Figure 7(a)} 
\end{center}
\end{minipage}

\begin{minipage}[t]{15cm}  
\epsfxsize=15cm
\epsfbox{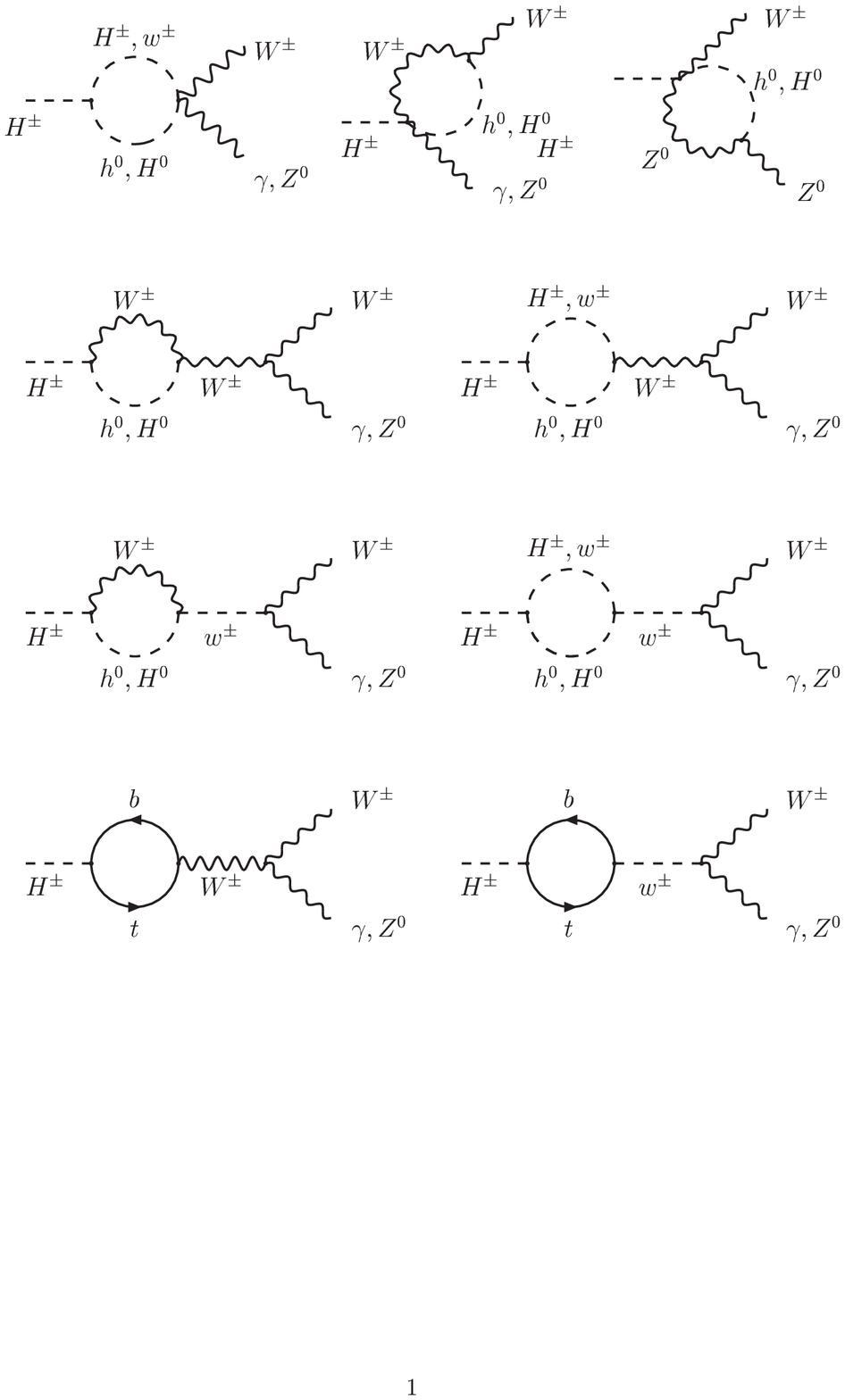}
\begin{center}
  {\bf Figure 7(b)} 
\end{center}
\end{minipage}

\begin{minipage}[t]{15cm}  
\epsfxsize=15cm
\epsfbox{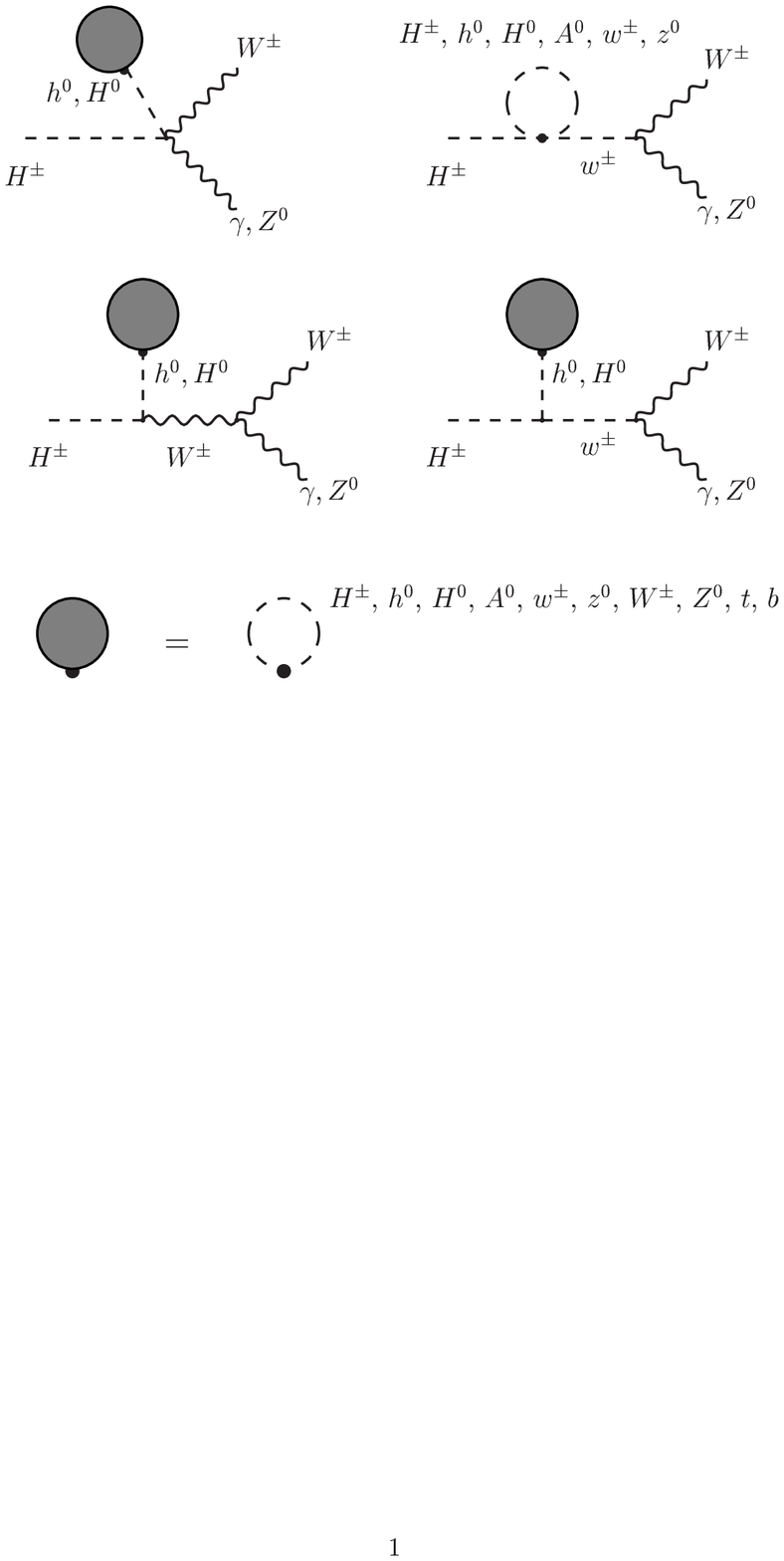}
\begin{center}
  {\bf Figure 7(c)} 
\end{center}
\end{minipage}


\begin{thebibliography}{99}
\bibitem{lc} {\it JLC-1},~KEK~Report~92-16~(1992); 
             {\it Physics and Technology of the Next Linear Collider}: 
                {\it a Report submitted to Snowmass 1996,} 
             BNL 52-502, FNAL-PUB-96/112, LBNL-PUB-5425, SLAC Report 485, 
             UCRL-ID-124160;  
             TESLA TECHNICAL DESIGN REPORT: Physics and Detector 
              ({\tt http://www.desy.de/~behnke/tdr/Welcome.html}). 
\bibitem{mp} A.~M{\'e}ndez and A.~Pomarol, 
                  \Journal{\NPB}{349}{369}{1991}.
\bibitem{cphi}   M.~Capdequi Peyran\`{e}re, H.E.~Haber and P.~Irulegui, 
                \Journal{\PRD}{44}{191}{1991}. 
\bibitem{kanemu} S.~Kanemura, \Journal{\PRD}{61}{095001}{2000}.
\bibitem{eehh} A.~Arhrib, M.~Capdequi Peyran\`{e}re and G.~Moultaka, 
               \Journal{\PLB}{341}{313}{1995};
               J.~Guasch, W.~Hollik and A.~Kraft, 
               Talk at the Vth workshop in the 2nd ECFA/DESY Study on Physics 
               and Detectors For a Linear Electron-Positron Collider, 
               Obernai (France) 16-19th Oct. 1999 ({\tt hep-ph/9911452}).
\bibitem{dec}  T.~Appelquist and J.~Carazzone, 
               \Journal{\PRD}{11}{2856}{1975}.
\bibitem{kko}    S.~Kanemura, T.~Kasai and Y.~Okada, 
                 \Journal{\PLB}{471}{182}{1999}.
\bibitem{nondec} P.~Ciafaloni and D.~Esprin, 
                 \Journal{\PRD}{56}{1752}{1997};
                 S.~Kanemura and H-A.~Tohyama,
                 \Journal{\PRD}{57}{2949}{1998}.  
\bibitem{uni} H.~H\"{u}ffel and G.~Pocsik, {\em Z. Phys.} C {\bf 8} (1981) 13; 
              J.~Maalampi, J.~Sirkka and I.~Vilja, 
              \Journal{\PLB}{265}{371}{1991};  
              S.~Kanemura, T.~Kubota and E.~Takasugi,
              \Journal{\PLB}{313}{155}{1993}.  
\bibitem{tri} 
  H.~Komatsu, \Journal {\PTP}{67}{1177}{1982};
  R. A.~Flores and M.~Sher, {\em Ann. Phys. (NY)} {\bf 148} (1983)  295;
  M.~Sher, {\it Phys. Rep.} {\bf 179} (1989) 273; 
  D.~Kominis and R.S.~Chivukula, \Journal {\PLB}{304}{152}{1993}; 
  S.~Nie and M.~Sher, \Journal {\PLB}{449}{89}{1999}. 
\bibitem{tanb} Y.~Grossman, \Journal{\NPB}{426}{355}{1994}.
\bibitem{gr}  A.K.~Grant, \Journal{\PRD}{51}{207}{1995};
              P.H.~Chankowski, M.~Krawczyk and J.~Zochowski, 
                          \Journal{\EPC}{11}{661}{1999}.     
\bibitem{bsg_ex} 
              CLEO Collaboration, CLEO CONF 98-17, ICHEP98 1011. 
\bibitem{bsg_mg} 
  M.~Ciuchini, G.~Degrassi, P.~Gambini and G.F.~Giudice,
   \Journal{\NPB}{527}{21}{1998}; 
  P.~Ciafaloni, A.~Romanino and A.~Strumia, 
   \Journal{\NPB}{524}{361}{1998};
  F.~Borzumati and G.~Greub, 
   \Journal{\PRD}{58}{074004}{1998}, 
   \Journal{\ibid} {59}{057501}{1999}; 
  T.M.~Aliev and E.O.~Iltan,
   \Journal{\PRD}{58}{095014}{1998}.
\bibitem{hhg} J.F.~Gunion, H.E.~Haber, G.~Kane and S.~Dawson,
              {\it The Higgs Hunter's Guide}, 
             (Addison-Wesley, New York, 1990).
\bibitem{brs} S.~Alam, K.~Hagiwara, S.~Kanemura, R.~Szalapski and 
              Y.~Umeda, \Journal{\NPB}{541}{50}{1999}. 
\bibitem{Hgamma} A.~Djouadi, V.~Driesen, W.~Hollik and 
                 J.~Rosiek, 
                 \Journal{\NPB}{491}{68}{1997}. 
\bibitem{reno} A. Dabelstein, \Journal{\NPB}{456}{25}{1995};
               M.~B\"{o}hm, W.~Hollik and H.~Spiesberger, 
              {\it Fortschr. Phys.} {\bf 34} (1986) 11.
\bibitem{mix-cap}
 M.~Capdequi~Peyran\`{e}re, \Journal{\IJMPA}{14}{429}{1999}.
\bibitem{eeav} A.G.~Akeroyd, A.~Arhrib and M.~Capdequi Peyran\'{e}re, 
                 \Journal{\MPLA}{14}{2093}{1999}. 
\bibitem{hhw} S.~Heinemeyer, W.~Hollik and G.~Weiglein,
             \Journal{\JHEP}{0006}{009}{2000}.
\bibitem{DELPI} DELPHI: LEPC talks, Presentations on Nov 9 1999, \\ 
                 {\tt 
              (http://delphiwww.cern.ch/\~offline/physics\_links/lepc.html)}
\bibitem{susytb} V.~Barger, M.S.~Berger and P.~Ohmann, 
                 \Journal{PRD}{47}{1993}{1093};
                 M.~Carena, M.~Olechowski, S.~Pokorski 
                 and C.E.M.~Wagner,  
                 \Journal{\NPB}{426}{1994}{269}. 
\bibitem{zhu}   S.H.~Zhu, ({\tt hep-ph/9901221}).
\bibitem{et} J.M.~Cornwall, D.N.~Levin and G.~Tiktopoulos, 
              \Journal{\PRL}{30}{1268}{1973}; 
              \Journal{\PRD}{10}{1145}{1974}; B.W.~Lee, 
             C.~Quigg and H.B.~Thacker, 
              \Journal{\PRD}{16}{1519}{1977}.
%
\bibitem{et2} M.S.~Chanowitz and M.K.~Gaillard, 
              \Journal{\NPB}{261}{379}{1985}; 
              H.~Veltman, \Journal{\PRD}{41}{2294}{1990};  
              J.~Bagger and C.~Schmidt, 
              \Journal{\PRD}{41}{264}{1990}; 
              H.-J. He, Y.-P. Kuang and X. Li, 
              \Journal{\PRL}{69}{2619}{1992}; 
              \Journal{\PRD}{49}{4842}{1994}. 
\bibitem{oda-pr} K.~Odagiri, {private communication}.
\bibitem{oda-top}  K.~Odagiri, \Journal{\PLB}{452}{327}{1999}.
\bibitem{achm} A.~Arhrib, M.~Capdequi Peyran\'{e}re, W. Hollik 
               and G. Moultaka, ({\tt hep-ph/9912527}).
\bibitem{pave} G.~Passarino and M.~Veltman, 
               \Journal{\NPB}{160}{151}{1979}.
\end{thebibliography}
\end{document}